\documentclass[final]{siamltex}

\usepackage{fullpage}
\usepackage{latexsym}
\usepackage{amsmath}

\usepackage{graphicx}
\usepackage{wrapfig}
\usepackage{subfigure}

\usepackage{amsmath}
\usepackage{amssymb}
\usepackage{amsbsy}

\newcommand{\vep}{\varepsilon}

\newcommand{\bT}{{\boldsymbol T}}

\newcommand{\fbb}{{\boldsymbol b}}

\newcommand{\bg}{{\boldsymbol g}}
\newcommand{\bq}{{\boldsymbol q}}

\newcommand{\fbf}{{\boldsymbol f}}

\newcommand{\bv}{{\boldsymbol v}}

\newcommand{\bx}{{\boldsymbol x}}
\newcommand{\by}{{\boldsymbol y}}
\newcommand{\bz}{{\boldsymbol z}}

\newcommand{\bX}{{\boldsymbol X}}

\newcommand{\bxi}{{\boldsymbol \xi}}

\def\div{\mathop{\rm div}\nolimits}

\numberwithin{equation}{section}

\title{Deconvolution closure for mesoscopic continuum models of\\
particle systems}
\author{Alexander Panchenko\thanks{Department of Mathematics, Washington State University, Pullman, WA 99164 ({\tt panchenko@math.wsu.edu}).}
\and
Lyudmyla L. Barannyk\thanks{Department of Mathematics University of Idaho, Moscow, ID 83844 ({\tt barannyk@uidaho.edu}). The work of this author was funded in part by Battelle Energy Alliance, LLC (BEA).}
\and
Kevin Cooper\thanks{Department of Mathematics, Washington State University, Pullman, WA 99164 ({\tt kcooper@math.wsu.edu}).}
}

\begin{document}

\maketitle

\bibliographystyle{siam}

\begin{abstract}
The paper introduces a general framework for derivation of continuum equations governing meso-scale dynamics of large particle systems.  The balance equations for spatial averages such as density, linear momentum, and energy were previously derived by a number of authors. These equations are not in closed form because the stress and the heat flux
cannot be evaluated without the knowledge of particle positions and velocities. We propose a closure method for approximating fluxes in terms of other meso-scale averages. The main idea is to rewrite the non-linear averages as linear convolutions that relate micro- and meso-scale dynamical functions. The convolutions can be approximately inverted using regularization methods developed for solving ill-posed problems. This yields closed form constitutive equations that can be evaluated without solving the underlying ODEs.  We test the method numerically on Fermi-Pasta-Ulam chains with two different potentials: the classical Lennard-Jones, and the purely repulsive potential used in granular materials modeling. The initial conditions incorporate velocity fluctuations on scales that are smaller than the size of the averaging window. The results show very good agreement between the exact stress and its closed form approximation.
\end{abstract}

\begin{keywords} 
FPU chain,  particle chain, oscillator chain, upscaling, model reduction, dimension reduction, closure
\end{keywords}


\begin{AMS}
82D25, 35B27, 35L75, 37Kxx, 70F10, 70Hxx, 74Q10, 82C21, 82C22
\end{AMS}


\section{Introduction}
Particle systems governed by Newton's ordinary differential equations (ODEs) are common in physics, engineering and computational biology.  Such systems could represent either classical molecular models, or discretizations of continuum mechanical partial differential equations (PDEs).  These ODE systems are difficult to simulate directly because of their large size and stiffness. Popular explicit solvers such as Verlet method require small time steps, which places severe restrictions on the length of the simulated time interval. This necessitates the development of new methods for reducing computational complexity.



Often, the ODE solutions are of secondary interest compared with various space-time averages. Examples of the latter are
density, velocity, stress, deformation gradient, and energy.
The averages can be always simulated directly, but a better option is to
formulate an approximate, continuum mechanical type model that describes the evolution of the averages. Such a model can be simulated at a lower cost than the underlying ODE system.

The theory of space-time averaging for particle systems was developed by several authors, starting with Irving and Kirkwood
\cite{Kirkwood} and Noll \cite{Noll}. Several decades later, Hardy \cite{Hardy} and Murdoch and Bedeaux \cite{mb}, \cite{mb96}, \cite{mb97}, \cite{murdoch07} developed the theory further and derived the governing balance equations.
The averaging in these works is done as follows. First, one selects the {\it primary averages} that would describe the meso-scopic state of the ODE system. These may be, for example, average density, velocity, deformation map, kinetic energy etc. Next, differentiating in time and using the ODEs, one obtains the governing balance equations that model meso-scale behavior of the particle system. The fluxes (or {\it secondary averages}) in these equations are given by explicit functionals of the ODE solutions. This establishes a connection between the fine scale ODE model, and the meso-scale PDE model.

While important for clarifying the relationship between micro- and meso-scale phenomena, results of this type do not provide a continuum model in the true sense of the word, because one still has to solve the ODEs to evaluate the fluxes. Therefore, the
balance equations in \cite{mb} are not in closed form.  In classical continuum mechanics, the constitutive equations express stress and heat flux in terms of velocity gradient, deformation gradient, and temperature.  In the above referenced theories, the average deformation and temperature are not sufficient for evaluating fluxes, because one still has to recover the positions and velocities of all particles. As a result, the complexity of the meso-scale models in \cite{mb} and \cite{Hardy} is about the same as the complexity of the ODE model.

In this paper, we propose a method for approximating fluxes in terms of the primary averages. The approximations have the same low complexity as the classical constitutive equations, but unlike these equations, our approximations combine explicit formulas and numerical algorithmic prescriptions.
From our point of view, any constitutive equation will be ultimately realized as a computational method. The accuracy and efficiency of this method are the main factors that determine the quality of the constitutive approximation. Therefore, instead of trying to construct short explicit equations, we sought a computational scheme that  (i) does not require solving the ODE system; (ii) clearly and consistently incorporates the micro-scale force equations; and (iii)  reproduces the exact fluxes with reasonable accuracy.

The stress in our constitutive equations depends on density and momentum in a non-local and non-linear manner. The non-locality makes our theory similar to the peri-dynamical formulation of continuum mechanics \cite{Lehoucq1}.
The difference between our approach and phenomenological peri-dynamics is that our method clearly links the micro-scale features of the particle dynamics and meso-scale constitutive equations. On the other hand, our work differs from the recent paper \cite{Lehoucq2} where the peri-dynamic stress and energy flux are given as exact functions of particle positions and velocities.  In our theory, the stress can be evaluated without solving the ODE system.

The main idea behind our approach is as follows. The primary averages are essentially non-linear integral operators acting on particle positions and velocities. These operators can be written as (linear) convolutions of the "window function" and certain functions of velocities and positions. In this way, each primary average is related to a function of the micro-scale dynamics. For example, density corresponds to the Jacobian of the inverse Lagrangian deformation map, and linear momentum corresponds to the product of this Jacobian and micro-scale velocity. The convolution operator is typically invertible, and thus micro-scale quantities can be, in principle, recovered from the averages. However,  the deconvolution problem is unstable (ill-posed) so that small perturbations of the averages can produce large perturbations in the recovered functions. Ill-possed problems are well studied in the literature (see e.g. \cite{Gr, Kirsch, Engl2, Morozov, Tikh1}) in both continuous and discrete settings.  A discrete version of the convolution integral equation is an ill-conditioned linear system. Such systems and related rank-deficient systems are treated in detail in the book \cite{Hansen}.  The strategy for solving ill-posed problems is to approximate the exact problem by a well-posed regularized problem depending upon a parameter. By varying this regularization parameter one obtains more accurate but less stable approximations. Loosely speaking, the effect of regularization is to smooth the exact solution and filter out higher frequency oscillations. The amount of smoothing and filtering depends of the choice of the method and the value of the regularization parameter, but some details of the exact solution are always lost. Despite this, it is always possible to produce a regularized deconvolution that is much closer to the exact micro-scale function than the corresponding average. The difference between the average and the regularized solution can be quite dramatic. The convolution, whose kernel is associated with the mesoscopic length scale, smears smaller details to such an extent that it is often impossible to recognize them by inspecting the graph of the average. A well chosen regularization performs a triage of length scales below the mesoscopic length: the smallest are filtered out, and the rest are recovered.

Let us now describe the main steps of the method.
\begin{enumerate}
\item[(i)] The starting point is a fine scale model: an ODE system of Newton's equations. We limit ourselves to the case of pairwise, short-range interaction forces that may depend on the relative positions and velocities. The system size $N$ is very large.
A typical interparticle distance is characterized by
a small parameter
\begin{equation}
\label{small-par}
\vep=N^{-1/d},
\end{equation}
where $d$ is the physical space dimension
(usually 1, 2,  or 3).
The particle masses and forces are scaled by $\vep$. The purpose of the scaling is to satisfy the following natural requirements. As $N\to\infty$, the total mass of the system should remain fixed, and the total particle energy should be either fixed,
or at least bounded independent of $N$.

The scaled ODE systems with increasing $N$ form a family of discrete models representing a {\it single} continuum system at different levels of micro-scale resolution. To prescribe initial conditions, we first fix the initial velocity interpolant. Then, given $N$, the initial particle velocities are generated by discretizing the interpolant on the uniform mesh of size $\vep L$. The initial positions are the nodes of the same mesh.
 \item[(ii)]Choose spatial mesoscale resolution parameter $\eta$. 
Fix a function $\psi$ with integral equal to one and compact support (non-compactly supported functions such as Gaussian are also possible). Then scale this function by a factor of $\eta$ and define the window function
\begin{equation}
\label{window}
\psi_\eta(\bx)=\eta^{-d}\psi\left(\frac{\bx}{\eta}\right).
\end{equation}
The window function is used to set up meso-scale averaging.
 \item[(iii)] Choose the primary mesoscopic averages (e.g. density, velocity, internal energy density, temperature). The collection of the primary averages represents the state of the particle system at the mesoscale.
\item[(iv)] Write down the exact balance equations for the primary variables.
Determine which fluxes in these equations require closure, and list all the microsopic quantities (e.g. interpolants of particle positions and velocities, Jacobians) that need to be reconstructed to achieve closure.
\item[(v)] Verify that the every required micro-scale
quantity can be (approximately) reconstructed from the chosen primary variables.  In each flux, replace the exact microscopic quantities with their
regularized deconvolution approximations, thereby obtaining approximations of the microscale quantities in terms of the primary averages.
\end{enumerate}

We tested the method numerically on one-dimensional Hamiltonian systems with short-range pair potentials. In the physics literature such systems are known as Fermi-Pasta-Ulam (FPU) chains. We consider two potentials: the classical Lennard-Jones, and another potential similar to the Hertz potential of granular acoustics.
Assuming that the meso-scale state of a system can be described by the density and linear momentum, we provide the corresponding balance equations and derive constitutive equations for the stress. The exact stress and exact primary averages are produced by the direct simulations with 10,000 particles. The approximate stresses are obtained by using deconvolution and then substituting into the formulas for the exact stress.  Then we compare the exact and approximate stresses rendered on the meso-scale mesh with 500 nodes. The results show that the approximation agrees very well with the exact stress.


The present article extends and improves the method introduced in \cite{PBG}.
In \cite{TPF}, some of the tools from \cite{PBG} are applied to the discrete models of fluids. In both papers, the suggested deconvolution algorithm was the classical Landweber iteration \cite{Fridman}, \cite{Land}.  Increasing the number of iterations $n$ increases accuracy but generally decreases stability. The zero-order approximation ($n=0$) consists of replacing micro-scale quantities with their averages.  This zero-order closure was studied in detail in \cite{PBG}. In \cite{TPF} we also used the first- and second-order approximations. Numerical experiments show that low-order closures work well when the initial velocity has small fluctuations, and the dynamics is nearly isothermal, meaning that the energy of velocity fluctuations is much smaller than the potential energy.

The Landweber iteration is simple and useful for modeling, but has a slow convergence rate (see \cite{Hansen}). For initial conditions with high fluctuations, a large number of iterations may be needed to achieve a reasonable accuracy.
In this work we use different techniques: regularization by discretization \cite{Kirsch} and truncated singular value decomposition (SVD) (see, e.g. \cite{Hansen}). Both methods are non-iterative. Regularization by discretization is straightforward: the integral is approximated by a numerical quadrature, and this eliminates accumulation of the spectrum to zero. In the truncated SVD method, the exact solution is represented in the basis of singular vectors. The regularized approximation is generated by discarding the components corresponding to the smallest singular values.  Using SVD yields additional computational savings, since the convolution kernel is dynamics-independent. The SVD of the kernel can be pre-computed and used repeatedly with different ODE systems.


Recently, a deconvolution approach was used in large eddy simulation (LES) of turbulence \cite{Adams-Stolz2}, \cite{Layton}, \cite{Layton06},  \cite{Layton07}.  In these works, deconvolution was used to approximate quadratic functions of velocity fluctuations by an operator acting on the average velocity. The present work differs from LES in several respects. The first difference is in the structure of the averaging operators. In LES, the average velocity depends linearly on the micro-scale velocity, while in our work this dependence is non-linear. This non-linearity makes it possible to handle the general ODE flows with non-constant Jacobians. The second difference is in the modeling. We provide the connection between Newtonian particle mechanics on the one hand, and continuum theories on the other hand. In LES, the objective is to simulate large scale features of flows governed by Navier-Stokes equations. The third difference is that the papers on LES do not systematically address ill-posedness, and do not make much use of the available results on ill-posed and inverse problems. The Gaussian kernel, one of the most popular in LES, is not the best in terms of stability of reconstruction, because the degree of ill-posedness depends on the smoothness of the kernel in the Sobolev scale: the smoother the kernel, the more unstable the reconstruction problem. For this reason, we use piecewise polynomial continuous kernels, which leads to a mildly ill-posed problem.

The paper is organized as follows. In Section 2 we describe a general multi-dimensional microscopic model and provide the exact balance equations for the averages, following \cite{mb}, \cite{murdoch07}.
In Section 3 we develop general multi-dimensional integral approximations of averages, and describe the use of regularization for approximate deconvolution. This section is the central section of the paper.
Section 4 contains the formulation of the scaled ODE equations of
FPU chains. In Section 5 we derive  closed form
balance equations of mass and momentum for such chains and provide the constitutive equations.
Section 6 contains the results of computational tests.  Finally, conclusions are given in Section 7.

\section{Microscale equations and mesoscale spatial averages}
\subsection{Scaled ODE problems}
We work with classical Newton
equations of point particle dynamics. The same equations may arise as discretization of the
momentum balance equation for continuum systems.
Consider a system containing $N\gg 1$ identical particles, denoted by $P_i$. The mass of each particle
is $\frac{M}{N}$, where $M$ is the total mass of the system. Suppose that during the
observation time $T$, $P_i$ remain inside a bounded domain $\Omega$ in ${\mathbb R}^d$,
where $d$ is the physical space dimension, usually $1, 2$ or $3$.
The positions $\bq_i(t)$ and velocities $\bv_i(t)$ of particles satisfy
 a system of ODEs
\begin{eqnarray}
\dot\bq_i&=&\bv_i,\label{one}\\
 \frac{M}{N}\dot\bv_i&=&\fbf_i+\fbf_i^{(ext)},\label{two}
\end{eqnarray}
subject to the initial conditions
\begin{equation}
\label{inits}
\bq_i(0)=\bx_i, \hspace*{1.0cm} \bv_i(0)=\bv^0_i.
\end{equation}
Here $\fbf_i^{(ext)}$ denotes external forces, such as gravity and confining forces. The interparticle forces $\fbf_i=\sum_j \fbf_{ij}$, where $\fbf_{ij}$ are pair interaction forces which depend on the relative positions and velocities of the respective particles.

We are interested in investigating asymptotic behavior of the system as $N\to \infty$. Thus it is convenient to introduce
a small parameter $\vep$ by (\ref{small-par})
characterizing a typical distance between neighboring particles. As $\vep$ approaches zero, the number of particles goes to infinity, and the distances between neighbors shrink. Consequently,
the
forces in (\ref{two}) should be properly scaled. The guiding principle for scaling is to make
the energy of the system bounded independent of $N$, as $N\to\infty$. In addition, the energy of the
initial conditions should be bounded uniformly in $N$.

As an example of scaling, consider  forces generated by a finite range
pair potential $U(\xi)$, where $\xi$ is the distance between the particles.
Suppose that each particle interacts with no more than a fixed number of neighbors. This implies
that there are about $N$ interacting pairs. If the system is sufficiently dense, and variations of
particle concentrations are not large, then a typical distance between interacting particles is on the order
$N^{-1/d}L=\vep L$. The resulting scaling
\begin{equation}
\label{force-scaled}
\fbf_{ij}
=-\frac{1}{N}\nabla_{\bq_i} U\left(\frac{|\bq_i-\bq_j|}{\vep}\right)=
-\frac{1}{\vep N}\frac{d}{d\xi} U(\xi)_{\left|_{\xi=|\bq_i-\bq_j|}\right.} \frac{\bq_i-\bq_j}{|\bq_i-\bq_j|}
\end{equation}
makes the potential energy of an isolated system bounded independent of $N$. Kinetic energy will be under control
provided the total energy of the initial conditions is bounded independent of $N$.
If exterior forces are present, they should be scaled as well.

\noindent
{\it Remark}. Superficially, the system (\ref{one}), (\ref{two}) looks similar to the parameter-dependent ODE systems studied
in numerous works on ODE time homogenization (see e.\,g. \cite{pavliotis-stuart} and references therein). In the problem under study,
$\vep$ depends on the {\it system dimension} $N$, while in the works on time-homogenization and ODE perturbation theory, the system size is usually fixed as $\vep\to 0$.

\subsection{Length scales}
We introduce the following length scales:\newline
\noindent
- macroscopic length scale $L={\rm diam}(\Omega)$;\newline
\noindent
- microscopic length scale
$\vep L$;\newline
\noindent
- mesoscopic length scale $\eta L$,\newline
\noindent
where
$0<\eta< 1$ is a parameter that characterizes spatial mesoscale resolution. This parameter is chosen based on the
desired accuracy, the computational cost requirements, and prior information about initial conditions and ODE
trajectories.

The computational domain $\Omega$ is subdivided into mesoscopic cubic cells $C_\beta$,
$\beta=1, 2,\ldots,  B$ with the side length on the order of $\eta L$.  The centers $\bx_\beta$ of $C_\beta$ are the nodes of
the meso-mesh. The number of unknowns in the mesoscopic system will be on the order of $B$. For computational efficiency,
one should have $B\ll N$. This does not mean that $\eta$ is close to one. In fact, it makes sense to keep $\eta$ as small as
possible in order to have an additional asymptotic control over the system behavior. Decreasing $\eta$ will in general
make computations more expensive.



\subsection{Averages and their evolution}
To define averages we first select a fast decreasing window function
$
\psi
$
satisfying
$
\int \psi(\bx) d\bx=1.
$
There are many possible choices of the window function. We assume that $\psi$ is a compactly supported, continuous, differentiable almost everywhere on the interior of its support, and non-negative. Next, define the window funciton $\psi_\eta$ by (\ref{window}).

Once the window function is chosen, one can generate averages of micro-scale dynamical functions (\cite{mb}, \cite{murdoch07}).
 The mesoscopic average density and momentum are given, respectively, by
\begin{equation}
\label{density}
\overline{\rho}^\eta(t, \bx)=\frac{M}{N}\sum_{i=1}^N \psi_\eta(\bx-\bq_i(t)),
\end{equation}
\begin{equation}
\label{mom}
\overline{\rho}^\eta \overline{\bv}^\eta(t, \bx)=\frac{M}{N}\sum \bv_i(t)\psi_\eta(\bx-\bq_i(t)).
\end{equation}
The meaning of the above definitions becomes clear if one considers $\psi=(c_d)^{-1} \chi(x)$, where $\chi$ is a characteristic function of the unit ball in ${\mathbb R}^d$, and $c_d$ is the volume of the unit ball. Then
$$
\overline{\rho}^\eta=\frac{1}{c_d \eta^d}\frac{M}{N} \sum \chi\left(\frac{\bx-\bq_i(t)}{\eta}\right).
$$
The sum in the right hand side gives the number of particles located within distance $\eta$ of $\bx$ at time $t$.
Multiplying by $M/N$ we get the total mass of these particles, and dividing by $c_d \eta^d$ (the volume of $\eta$-ball) gives
the usual particle density.

Differentiating (\ref{density}), (\ref{mom}) in $t$, and using the ODEs (\ref{one}), (\ref{two}) one can obtain \cite{mb}
exact mesoscopic balance equations for the primary variables. For example, for an isolated system with ($\fbf_i^{(ext)}=0$), mass conservation and momentum balance equations take the form:
\begin{equation}
\label{mass-balance}
\partial_t \overline{\rho}^\eta+\div (\rho^\eta\overline{\bv}^\eta)=0,
\end{equation}
\begin{equation}
\label{m-balance}
\partial_t(\overline{\rho}^\eta\overline{\bv}^\eta)+\div\left(\overline{\rho}^\eta\overline{\bv}^\eta\otimes
\overline{\bv}^\eta\right) - \div\bT^\eta=0.
\end{equation}
The stress
$\bT^\eta=\bT^\eta_{(c)}+\bT^\eta_{(int)}$ \cite{murdoch07}, where
\begin{equation}
\label{m-stress-c}
\bT^\eta_{(c)}(t, \bx)=-\sum m_i(\bv_i-\overline{\bv}^\eta(t, \bx, ))\otimes (\bv_i-\overline{\bv}^\eta(\bx, t))\psi_\eta(\bx-\bq_i)
\end{equation}
is the {\it convective stress}, and
\begin{equation}
\label{m-stress-int}
\bT^\eta(t, \bx)_{(int)}=
\sum_{(i, j)}\fbf_{ij}\otimes (\bq_j-\bq_i)\int_0^1 \psi_\eta\left(s(\bx-\bq_j)+(1-s)(\bx-\bq_i)\right)ds
\end{equation}
is the {\it interaction stress}. The summation in (\ref{m-stress-int}) is over all pairs of particles $(i, j)$ that interact with each other.

Discretizing balance equations on the mesoscopic mesh yields a discrete system of equations, called the {\it meso-system}, written for mesh values
of $\overline{\rho}^\eta_\beta, (\overline{\rho}^\eta\overline{\bv}^\eta)_\beta$ and $\bT^\eta_\beta$.
The dimension of the meso-system is much smaller than the dimension of the original ODE problem. However, at this stage we still have no computational savings, since the meso-system is {\it not closed}. This means that mesoscopic fluxes such as
(\ref{m-stress-c}), (\ref{m-stress-int}) are expressed as functions of the microscopic positions and velocities. To find these
positions and velocities, one has to solve the original microscale system (\ref{one}), (\ref{two}).
To achieve computational savings we need to replace exact fluxes with approximations that involve only mesoscale quantities.
We refer to the procedure of generating such approximations as a {\bf closure method}.
This closure-based approach has much in common with continuum mechanics.
The important difference is that the focus is on computing, rather than continuum mechanical style
modeling of constitutive equations.
\section{Closure via regularized deconvolutions}
\subsection{Outline}
Our approach is based on a simple idea: the integral approximations of primary averages (such as density and velocity) are related to the corresponding microscopic quantities via convolution with the kernel $\psi_\eta$.  Therefore, given primary variables we can (approximately) recover the microscopic positions and velocities by numerically inverting convolution operators. The results are inserted into equations for secondary averages (or fluxes), such as stress in the momentum balance. This yields closed form balance equations that can be simulated efficiently on the mesoscopic mesh.
\subsection{Integral approximation of discrete averages}
\label{sect:int-appr}
To exploit the special structure of primary averages, it is convenient to approximate sums such as
\begin{equation}
\label{gen-av}
\overline{g}^\eta=\frac{1}{N}
\sum_{j=1}^Ng(\bv_j, \bq_j) \psi_\eta(\bx-\bq_j)=\frac{1}{|\Omega|}\frac{|\Omega|}{N}
\sum_{j=1}^Ng(\bv_j, \bq_j) \psi_\eta(\bx-\bq_j)
\end{equation}
by integrals. The sum in (\ref{gen-av}) resembles a Riemann sum for $|\Omega|^{-1}g \psi_\eta(\bx-\cdot)$, where
$\Omega$ is partitioned into $N$ cells of volume $\frac{|\Omega|}{N}$, with one particle located inside of each cell. However, because of the motion of particles, (\ref{gen-av}) is not in general a Riemann sum. Indeed, to interpret this sum correctly, one must exhibit a partition of $\Omega$ into cells of equal volume where each cell contains exactly one particle. Such a partition may not exist. Indeed, in one dimension, the domain is an interval, say $(0, L)$ and the cells are intervals of length $L/N$.
Thus, if the closest neighbors of a given particles are less than $L/N$ apart, than the desired partition does not exist.
For two- and three-dimensional domains, it may be possible to use more general partitions, but for particles that are spaced non-uniformly, the shapes of these cells may be quite far from slightly deformed rectangles, which would make it difficult to estimate the accuracy of the resulting integral approximation.

A more systematic way to generate integral approximations is to use a microscopic flow map interpolant and the associated Jacobian describing local volume changes. Let
$\tilde{\bq}(t, \bX), \tilde{\bv}(t, \tilde{\bq})$ be suitable position and velocity interpolants, associated with the system
(\ref{one}), (\ref{two}). At $t=0$ these interpolants satisfy
$$
\tilde{\bq}(0, \bX_j)=\bq_j^0,~~~~~~~~~~\tilde{\bv}(0, \tilde{\bq}(0, \bX_j))={\bv}^0_j,
$$
where $\bX_j$, $j=1, 2, \ldots, N$ are points of $\vep$-periodic rectangular lattice in $\Omega$.
At other times,
$$
\tilde{\bq}(t, \bX_j)=\bq_j(t), ~~~~~~~\tilde{\bv}(t, \tilde{\bq}(t, \bX_j))=\bv_j(t).
$$

Then we can rewrite (\ref{gen-av}) as
\begin{equation}
\label{gen-av2}
\overline{g}^\eta=\frac{1}{|\Omega|}
\sum_{j=1}^N \frac{|\Omega|}{N} g
\left(
\tilde{\bv}
\left(t, \tilde{\bq}(t, \bX_j)
\right),
\tilde{\bq}(t, \bX_j
\right)
\psi_\eta(\bx-\tilde{\bq}(t, \bX_j)),
\end{equation}
where $|\Omega|$ denotes the volume (Lebesgue measure) of $\Omega$.
Eq. (\ref{gen-av2})  is a Riemann sum generated by partitioning $\Omega$ into $N$ cells of volume $|\Omega|/N$
centered at $\bX_j$. This yields
\begin{equation}
\label{int1}
\overline{g}^\eta =\frac{1}{|\Omega|} \int_\Omega g
\left(
\tilde{\bv}(t, \tilde{\bq}(t, \bX)), \tilde{\bq}(t, \bX)
\right)
\psi_\eta (\bx-\tilde{\bq}(t, \bX)) d\bX,
\end{equation}
up to discretization error.
Now suppose that the map $\tilde{\bq}(\cdot, \bX)$ is invertible for each $t$, that is $\bX=\tilde{\bq}^{-1}(t, \tilde{\bq})$. Changing the variables in the integral $\by=\tilde{\bq}(t, \bX)$ we obtain a generic integral approximation
\begin{equation}
\label{int2}
\overline{g}^\eta= \frac{1}{|\Omega|} \int_\Omega g\left(\tilde{\bv}(t, \by), \by
\right) \psi_\eta (\bx-\by) J(t, \by)~d\by,
\end{equation}
where
\begin{equation}
\label{J}
J=|\det \nabla \tilde{\bq}^{-1}|,
\end{equation}
up to discretization error.
For reader's convenience, we list the integral approximations of the average density and momentum.
\begin{eqnarray}
\label{int-density}
\overline{\rho}^\eta(t, \bx)& = &\frac{M}{N}\sum_{i=1}^N \psi_\eta(\bx-\bq_i(t))\\
& = & \frac{M}{|\Omega|}\int_\Omega \psi_\eta (\bx-\tilde{\bq}(t, \bX)) d\bX\nonumber\\
& = & \frac{M}{|\Omega|}\int_\Omega \psi_\eta (\bx-\by) J(t, \by) d\by.\nonumber
\end{eqnarray}
\begin{eqnarray}
\label{int-mom}
\overline{\rho}^\eta \overline{\bv}^\eta(t, \bx)& = &\frac{M}{N}\sum_{i=1}^N \bv_i(t)\psi_\eta(\bx-\bq_i(t))\\
& = & \frac{M}{|\Omega|}\int_\Omega \tilde{\bv}(t, \tilde{\bq}(t, \bX))\psi_\eta (\bx-\tilde{\bq}(t, \bX)) d\bX\nonumber\\
& = & \frac{M}{|\Omega|}\int_\Omega \psi_\eta (\bx-\by) \tilde{\bv}(t, \by) J(t, \by) d\by.\nonumber
\end{eqnarray}
Note the linear convolution structure of the $\by$-integrals in (\ref{int-density}) and (\ref{int-mom}).
It is also worth noting that these equalities are exact if the interpolants are piecewise linear. In that case, the discrete sums are exact integral quadratures.
\subsection{Regularized deconvolutions}
\subsubsection{General considerations}
Define an operator $R_\eta$ by
$$
R_\eta[f](\bx)=\int \psi_\eta(\bx-\by) f(\by) d\by.
$$
To simplify exposition, suppose that $R_\eta$ is injective.
In that case, there exists the single-valued
inverse operator $R^{-1}_\eta$, that we call the {\it deconvolution operator}. Since
$R_\eta$ is compact in $L^2(\Omega)$, the inverse operator is unbounded.  Therefore, small perturbations of the right hand side can lead to large perturbations in the computed solution.
Reconstructing $f$ from the knowledge of $R_\eta[f]$) is a classical example of an unstable ill-posed problem.
Such problems are well investigated both analytically and numerically (see, e. g.  \cite{Gr, Hansen, Kirsch, Morozov, Tikh1, Engl2}). Many solution techniques are currently available: Tikhonov regularization, iterative methods, reproducing kernel methods, the maximum entropy method, the dynamical system approach and others.
In the sequel we use the notation
$Q_\eta$ for a regularized approximation of the exact inverse
operator.

Recently, the classical Landweber iterative deconvolution method \cite{Fridman}, \cite{Land}  has attracted attention as a means to achieve sub-filter scale resolution in large eddy simulation of turbulence \cite{Adams-Stolz2}, \cite{Layton}.
In the simplest version of the method, approximations $g_n$ to the solution of the
operator equation
\begin{equation}
\label{op-eq}
R_\eta[g]=\overline{g}^\eta
\end{equation}
are generated by the formula
\begin{equation}
\label{seq1}
g_n=\sum_{k=0}^n (I-R_\eta)^n \overline{g}^\eta, \;\;\;\; g_0=\overline{g}^\eta.
\end{equation}
The number $n$ of iterations plays the role of regularization parameter.
In (\ref{seq1}), $I$ denotes the identity operator.

Another classical method is Tikhonov regularization \cite{Tikh1}, where the solution of (\ref{op-eq}) is approximated by
$g_\alpha$ that solves
\begin{equation}
\label{Tikhonov}
R_\eta[g_\alpha]+\alpha C[g_\alpha]=\overline{g}^\eta.
\end{equation}
Here $\alpha>0$ is a regularization parameter, and $C$ is a stabilizing operator. In practice, $C$ can be an identity, or a
a suitable differential operator such as Laplacian.


\subsubsection{Computational implementation of deconvolution}

The discrete version of the integral equation (\ref{op-eq}) is a linear system
\begin{equation}
\label{lin}
A\bg=\overline{\bg}.
\end{equation}
The solution $\bg$ is a discrete micro-scale quantity to be reconstructed, and $\overline{\bg}$ is the discretization of the corresponding average.
To achieve computational efficiency, it is natural to resolve the average (a meso-scale quantity) on a coarse mesh with the size tied to the meso-scale. The solution (a microscopic quantity) could be rendered on the fine mesh with size $\vep L$. This choice of meshes seems to be the most natural for balancing cost and accuracy.

Other mesh combinations can be chosen as well.  The least expensive option is to coarsen the discretization of $\bg$ and solve (\ref{lin}) entirely on a coarse mesh. In that case, the matrix $A$ is square, and its dimension is determined by the chosen number of the coarse mesh points. The operation count of deconvolution becomes independent of $N$, but this efficiency comes at the price of introducing too much error.  Numerical experiments produced significant artifacts, so that the meso-scale details of $\bg$ could not be reliably reconstructed.

Another possibility is to use a fine mesh for both $\bg$ and $\overline{\bg}$. In that case the average would have to be interpolated from the coarse to the fine mesh. The computational cost of deconvolution scales as $O(N^2)$, but this dos not  necessarily improve the reconstruction quality, since using a finer discretization increases the condition number of $A$.

Ultimately, we chose the more natural two-mesh discretization, whereby $\bg$ is an $N$-vector, $\overline{\bg}$ is a $D$-vector, with $B< D\ll N$. Recall that $B$ is the number of the mesh nodes associated with the mesh size $\eta L$. Choosing a finer coarse mesh with $D$ nodes enabled us to better resolve the details
of size smaller than $\eta L$. Such details may be completely smeared by averaging.  In that sense, the length scale associated with $D$ can be called a {\it sub-filter scale}.


Working with two different meshes, as opposed to using the same mesh, introduces several
difficulties. The matrix $A$ in that case is rectangular, and the system (\ref{lin}) is under-determined.
Therefore, (\ref{lin}) has many solutions, even when the original integral equation (\ref{op-eq}) is  uniquely solvable.
The general solution is a sum of the particular solution $\bg^+$ orthogonal to the null space of $A$ and an arbitrary vector from the null space. In the absence of a priori information on the structure of the null space, it is natural to use $\bg^+$. Thus we set
$$
\bg^+ = A^T \bz,
$$
where $\bz$ is a $D$-vector to be determined. Assuming that $A$ has full rank and $A A^T$ is invertible (invertibility depends only on the choice of the window function $\psi$ and can be verified prior to running any dynamic simulations),
we can set
$
\bz=(A A^T)^{-1} \overline{\bg},
$
and
\begin{equation}
\label{exact-plus}
\bg^+=  A^T (A A^T)^{-1} \overline{\bg}.
\end{equation}
It is easy to check that $\bg^+$ is orthogonal to the null space of $A$. In (\ref{exact-plus}), $(AA^T)^{-1}$ denotes either the exact inverse, or a suitable regularized approximation.
Typically, singular vectors associated with smaller singular values of $A$ oscillate with higher frequency than the vectors associated with larger singular values. If this holds, then the solution component along the null-space of $A$ is highly oscillatory, while the component orthogonal to the null space is relatively smooth. Therefore, by using $\bg^+$ and not some other solution, we incorporate additional filtering. This can be useful for taming noise.

\begin{figure}[h] \begin{center}
\includegraphics[height=1.5in,angle=0]{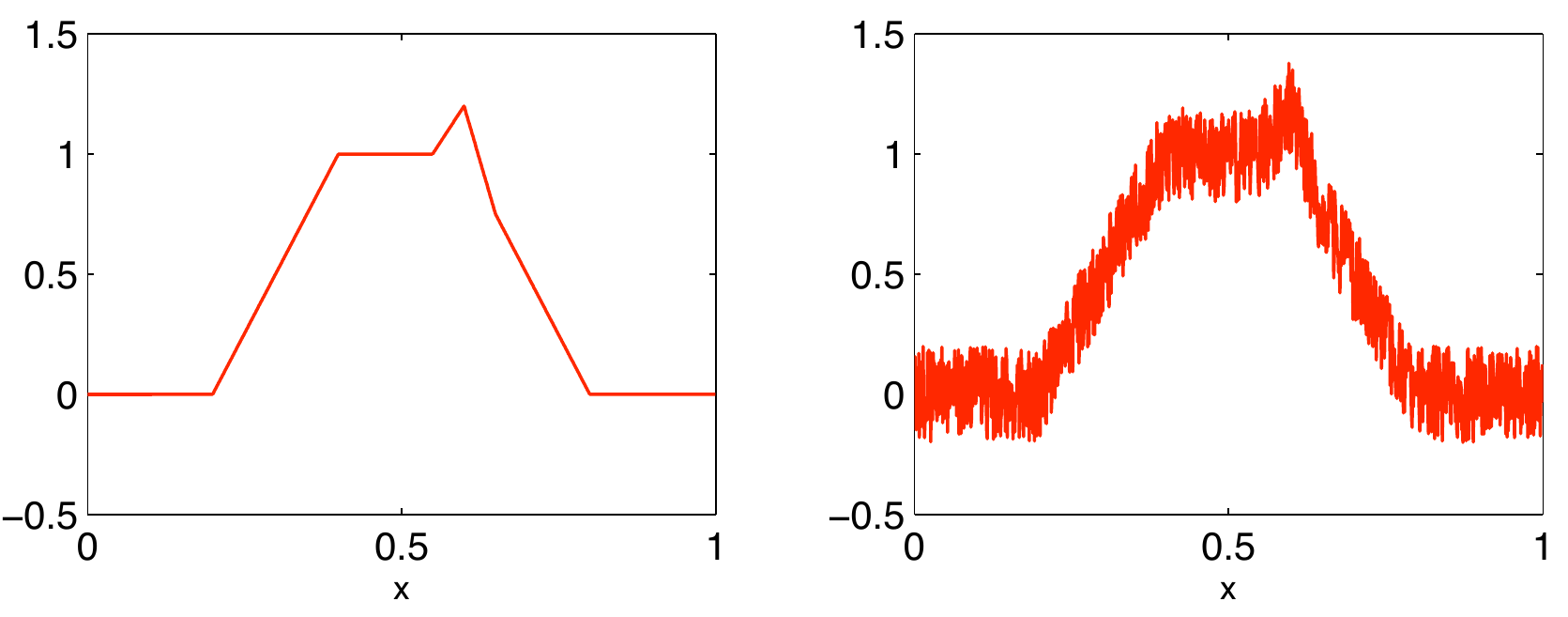}
\end{center}
\caption{Left panel: meso-scale and sub-filter scale features; right panel: exact solution with a uniformly distributed noise added}
\label{fig-deconv-one}
\end{figure}

\begin{figure}[h] \begin{center}
\includegraphics[height=1.5in,angle=0]{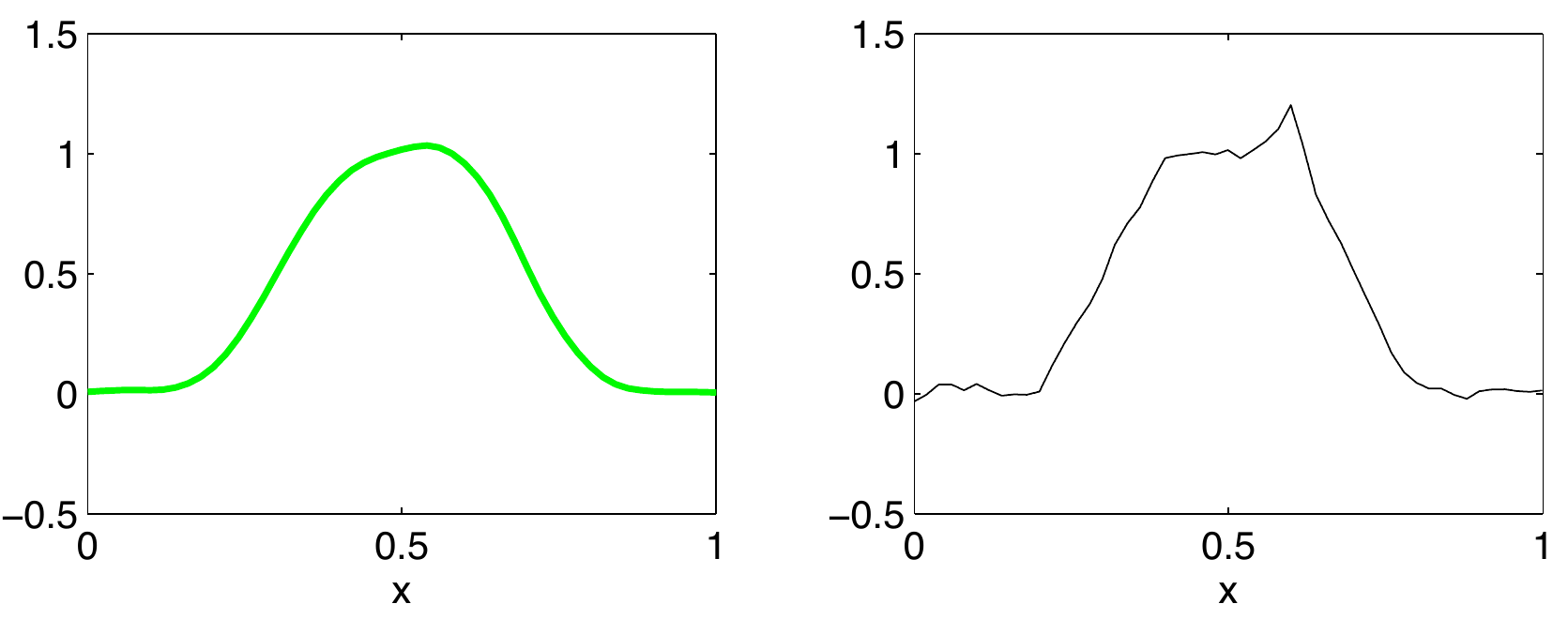}
\end{center}
\caption{Left panel: the average (right hand side of the integral equation); right panel: reconstructed approximate solution}
\label{fig-deconv-two}
\end{figure}

An example of deconvolution is shown in Figs. \ref{fig-deconv-one} and \ref{fig-deconv-two}. In the first of these figures we show the exact solution $\bg$. It is constructed by first choosing a profile to be reconstructed (left panel), and then adding noise (right panel). The graph in the left panel contains a meso-scale feature (trapezoidal impulse in the center), and a sub-filter scale feature (smaller triangular impulse on top of the trapezoid). The noise contains multiple features on smaller length scales.

The right hand side vector $\overline{\bg}$, computed by applying $A$ to $\bg$ in the right panel of Fig. \ref{fig-deconv-one}, is shown in Fig. \ref{fig-deconv-two}. All features except the largest appear to be smeared. The right panel in Fig. \ref{fig-deconv-two} shows the reconstruction computed using (\ref{exact-plus}). The noise is largely filtered out but the sub-filter feature is clearly visible.
%
%

%
%
%

Many regularization methods (Tikhonov, Landweber, truncated SVD) can be conveniently written in terms of  SVD and spectral filter functions (see e.g. \cite{Kirsch}). This is useful for both discrete and continuous ill-posed problems \cite{Hansen}. Since both $R_\eta$ and $A$ are independent of the microscopic dynamics, the SVD can be pre-computed and used with different ODE systems.

Suppose that $\text{rank}(A)=D$.
Let $\sigma_j,  j=1, \ldots, D$ denote the non-zero singular values of $A$, and
$\bxi_j\in {\bf R}^D$, $\widehat \bxi_j \in {\bf R}^N$ be the corresponding singular vectors. By the standard properties of SVD,
\begin{equation}
\label{svd}
A \widehat\bxi_j=\sigma_j \bxi_j,~~~~~ A^T \bxi_j=\sigma_j \widehat\bxi_j.
\end{equation}
Because the continuum problem (\ref{op-eq}) is ill-posed, the singular values are spaced without gaps, and the condition number of $A$ is large. Recall that condition number can be expressed as the ratio of the largest and smallest singular values. In our case, the largest singular value of $A$ is close to one. Therefore, the condition number is approximately equal to the reciprocal of the smallest singular value.
The choice of the solution method depends on the condition number and the relative level of noise in $\overline{\bg}$ and $A$.  The guiding principle it to produce an approximation that would be as close as possible to (\ref{exact-plus}), without incurring instability. Discretization itself is a mild form of regularization of the original integral equation. Indeed, in the discretized problem, the smallest non-zero singular value is a finite distance away from zero, while in the continuum case zero is an accumulation point of the spectrum. Consequently, if the error in $\overline{\bg}$ and the condition number of $A$ are small enough, one can use (\ref{exact-plus}) with no additional regularization.
In our case, the integral approximations of the averages are exact when the interpolants are suitably chosen. The only numerical error in the right hand side is the round-off error. Therefore, exact inversion will work when the condition number of $A$ is much smaller than the reciprocal of the machine precision. In practice this means that condition numbers smaller than about $10^8$ can be safely handled in this way. For larger condition numbers, exact inversion in (\ref{exact-plus}) may have to be replaced by a suitable regularized approximation.


First we describe the SVD-based implementation of the exact solution formula (\ref{exact-plus}).
Write
\begin{equation}
\label{diag}
\overline{\bg}=\sum_{j=1}^D g_j \bxi_j, \hspace*{1cm}\bg^+=A^T \left(\sum_{j=1}^D z_j \bxi_j\right)=\sum_{j=1}^D \sigma_j z_j \widehat\bxi_j,
\end{equation}
where the coefficients $z_j$ have to be determined.
To obtain the last equality, (\ref{svd}) was used. Substituting (\ref{diag}) into (\ref{lin}) and using orthogonality of $\bxi_j$ we deduce
\begin{equation}
\label{x-plus}
z_j=\frac{g_j}{\sigma_j},
\end{equation}
which represents (\ref{exact-plus}) in the basis consisting of singular vectors.

As was explained earlier, formula (\ref{x-plus}) may be used when condition number of $A$ is much smaller than the reciprocal of the noise level in the right hand side.
Therefore, (\ref{x-plus}) becomes unstable when the SVD contains singular values comparable to the machine precision. In that case, as an additional regularization we use the truncated SVD \cite{Hansen}. In this method, the components corresponding to the smallest singular values are discarded. The regularized solution is computed by the formula
\begin{equation}
\label{reg-plus}
\bg^+_{(r)}=\sum_{j=1}^D \frac{\phi(\sigma_j)}{\sigma_j}g_j \widehat \bxi_j,
\end{equation}
where the filter function $\phi$ is defined as follows.
\begin{equation}
\label{filter}
\phi(\sigma_j)=
\left\{
\begin{array}{cc}
1 & {\rm if}\; \sigma_j \geq \sigma^*\\
0 & {\rm if} \;\sigma_j < \sigma^*. \\
\end{array}
\right.
\end{equation}
In the above equation $\sigma^*$ is a cut-off value (equal to the machine precision in the present case).

\section{FPU chain equations}
In this section, the general method outlined above is detailed in the case of one-dimensional Hamiltonian chain of oscillators that consists of $N$ identical particles. The domain $\Omega$ is an interval $(0, L)$.
Particle positions, denoted by $q_j=q_j(t)$, $j=1,\ldots, N$, satisfy
\[
0<q_1<q_2<\ldots<q_N<L
\]
at all times, i.e. the particles cannot occupy the same position or jump over each other. Next, define a small
parameter
$$
\vep=\frac{1}{N},
$$
and microscale step size
\begin{equation}
\label{h}
h=\frac{L}{N}.
\end{equation}

The interparticle forces
\begin{equation}
\label{micro-force}
f_{jk}=
-\frac{q_j-q_k}{|q_j-q_k|}
U^{\prime}\left(\frac{|q_j-q_k|}{\vep}\right)
\end{equation}
are defined by a finite range potential $U$.

Each particle has mass $m=M/N=M\vep$, where $M$ is the total mass of the system.  Particles have velocities denoted by $v_j$, $j=1,\ldots,N$.
Writing the second Newton's law as a system of first order equations yields the scaled microscale ODE system
\begin{equation}
\label{ode-1}
\begin{array}{l}
\dot{q_j}=v_j,\hspace*{1cm} \vep M\dot{v_j}=f_j, \;\;\; j=1,\dots,N
\end{array}
\end{equation}
subject to the initial conditions
\begin{equation}
\label{ode-2}
q_j(0)=q_j^0, \hspace*{1cm} v_j(0)=v_j^0.
\end{equation}
\section{Integral approximation of stresses for particle chains. Mesoscopic continuum equations}
In the one-dimensional case stress is a scalar quantity, and (\ref{m-stress-c}), (\ref{m-stress-int}) reduce to, respectively,
\begin{equation}
\label{c1}
T^\eta_{(c)}(t, x)=-\sum_{j=1}^N \frac{M}{N}
(
v_j-\overline{v}^\eta(t, x)
)^2
\psi_\eta (x-q_j),
\end{equation}
and
\begin{equation}
\label{c2}
T^\eta_{(int)}(t, x)=
\sum_{j=1}^{N-1} f_{j, j+1} (q_{j+1}-q_j)\int_0^1 \psi_\eta(x-s q_{j+1}-(1-s)q_j)ds.
\end{equation}
The sum in (\ref{c2}) is simplified compared to the general expression (\ref{m-stress-int}), since we have exactly $N-1$ interacting pairs of particles.

To obtain integral approximations of stresses, we define interpolants $\tilde q, \tilde v$, as in Sect. \ref{sect:int-appr}.
Repeating the calculations we get
\begin{equation}
\label{c3}
T^\eta_{(c)}(t, x)=
-\frac{M}{L} \int_0^L \left(
\tilde v(t, y)-\overline{v}^\eta(t, x)
\right)^2
\psi_\eta(x-y) J(t, y) dy.
\end{equation}

\noindent
{\it Remark}. Many equalities in the paper, including (\ref{c3}) hold up to a discretization error. To simplify presentation, we do not mention this in the sequel when discrete sums are approximated by integrals.

The interaction stress can be rewritten as
\begin{equation}
\label{c4}
T^\eta_{(int)}(t, x)=-
\frac{N-1}{N}
\sum_{j=1}^{N-1}\frac{L}{N-1} U^\prime
\left(
\frac{q_{j+1}-q_j}{h}L
\right)
 \frac{q_{j+1}-q_j}{h}\int_0^1 \psi_\eta(x-s q_{j+1}-(1-s)q_j)ds.
\end{equation}
Next we approximate
\begin{equation}
\label{jacobian-discrete}
(q_{j+1}-q_j)/h\approx \tilde q^\prime(t, X)=\frac{1}{(\tilde q^{-1})^\prime(t, \tilde q(t, X))}=\frac{1}{J(t, \tilde q(t, X))}.
\end{equation}
This approximation is in fact exact, provided the interpolant is chosen to be piecewise linear. Note also that
this equation is a special feature of one-dimensional dynamics, where the derivative (deformation gradient) can be identified with the Jacobian of the deformation map. In higher dimensions, this no longer holds.
Inserting this into (\ref{c4}), replacing Riemann sum with an integral and changing variable of integration
as in Sect. \ref{sect:int-appr}, we obtain the integral approximation
\begin{equation}
\label{c5}
T^\eta_{(int)}(t, x)=-\frac{N-1}{N}
\int_0^L U^\prime
\left(
\frac{L}{J(t, y)}
\right)
\int_0^1 \psi_\eta
\left(
x-y-\frac{sh}{J(t, y)}
\right)
ds~
dy.
\end{equation}
Equations (\ref{c3}), (\ref{c5}) contain two microscale quantities: $J$ and $\tilde v$. Approximating sums in
the definitions of the primary averages (\ref{density}), (\ref{mom}) by integrals we see that $\overline{\rho}^\eta$ and $\overline{v}^\eta$ are obtained by applying the convolution operator $R_\eta$ to, respectively
$J$ and $J \tilde v$:
\begin{equation}
\label{c6}
\overline{\rho}^
\eta=\frac{M}{L} R_\eta [J],\hspace*{1.5cm}\overline{\rho}^
\eta\overline{v}^\eta=\frac{M}{L} R_\eta [J \tilde v].
\end{equation}

Recall that $Q_{\eta}$ denotes a regularizing approximation to the exact inverse operator $R_\eta^{-1}$.
Applying $Q_\eta$ in (\ref{c6}) yields integral approximations
\begin{equation}
\label{c7}
J\approx \frac{L}{M} Q_\eta [\overline{\rho}^
\eta], \hspace*{1.5cm}~~~~~~~~~ ~~~~~~\tilde v\approx \frac{Q_\eta[\overline{\rho}^
\eta \overline{v}^\eta]}{Q_\eta [\overline{\rho}^
\eta]}.
\end{equation}
Inserting these approximation into the integral formulas for the stress yields closed form mesoscopic continuum equations
\begin{eqnarray}
\partial_t \overline{\rho}^\eta+\partial_x (\overline{\rho}^\eta\overline{v}^\eta) &=& 0,\label{c8}\\
\partial_t( \overline{\rho}^\eta\overline{v}^\eta)+\partial_x
\left(
\overline{\rho}^\eta(\overline{v}^\eta)^2
\right)
-\partial_x (\overline{T}^\eta_{(c)}+\overline{T}^\eta_{(int)})& =& 0,\label{c9}
\end{eqnarray}
where $\overline{T}^\eta_{(c)}, \overline{T}^\eta_{(int)}$ are given by
\begin{equation}
\label{c10}
\overline{T}^\eta_{(c)}=
- \int_0^L \left(
\frac{Q_\eta[\overline{\rho}^
\eta \overline{v}^\eta]}{Q_\eta [\overline{\rho}^
\eta]}\left(t, y\right)-\overline{v}^\eta(t, x)
\right)^2
\psi_\eta(x-y) (t, y) Q_\eta [\overline{\rho}^
\eta](t, y)dy,
\end{equation}
\begin{equation}
\label{c11}
\overline{T}^\eta_{(int)}=-\frac{N-1}{N}
\int_0^L U^\prime
\left(
\frac{M}{Q_\eta[\overline\rho^\eta](t, y)}
\right)
\int_0^1 \psi_\eta
\left(
x-y-\vep\frac{Ms}{Q_\eta[\overline\rho^\eta](t, y)}
\right)
ds~
dy.
\end{equation}
 The choice of the operator $Q_\eta$ is not unique: it depends on the specifics of the regularization method and the chosen value of the regularization parameter. For classical regularization schemes such as Landweber, Tikhonov, and truncated SVD, this operator is a convolution with
the kernel ${\mathcal Q}(x)$ that can be described explicitly in terms of the singular values and singular vectors of $R_\eta$ (see \cite{Kirsch} for details).

In \cite{PBG} we studied a special case of (\ref{c7}) corresponding to the Landweber approximation (\ref{seq1}) with $n=0$. In that case, called zero-order closure, $Q$ is the identity operator. This means that
\begin{equation}
\label{zero-order}
J\approx J_0=\frac{L}{M} \overline{\rho}^\eta, \hspace*{1.3cm} \tilde v\approx \tilde v_0=\overline{v}^\eta.
\end{equation}

\section{Numerical experiments} \label{NumExper}
%
%
\subsection{Lennard-Jones chain} \label{LJchain}
\begin{figure}[h] \begin{center}
\includegraphics[height=1.5in,angle=0]{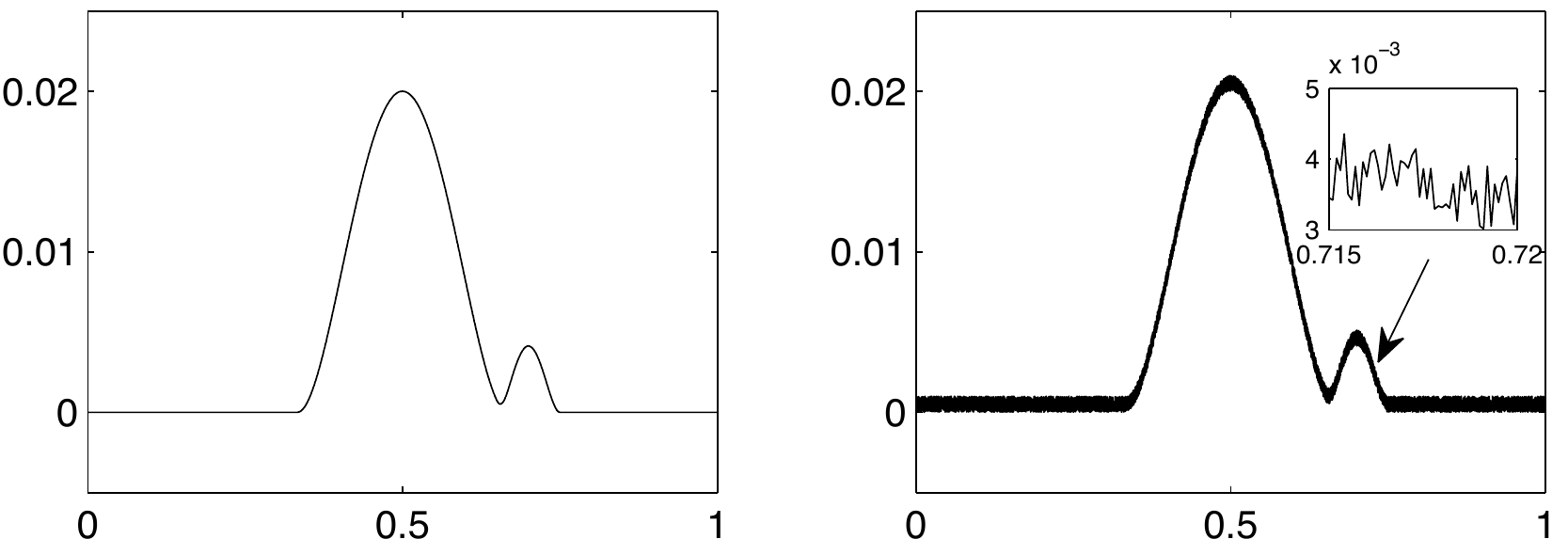}
\end{center}
\caption{Left panel: deterministic initial velocity; right panel: noisy initial velocity }
\label{fig_ICs}
\end{figure}

In this example, we simulate a chain of particles interacting with Lennard-Jones potential plotted in left panel of Fig. \ref{FLB_Hertz_potential} and defined in (\ref{LJ_def}) in the Appendix B. The initial positions are equally spaced with $q_j^0=(j-1/2)h$, $j=1,\ldots,N$.
We consider two different sets of initial velocities shown in Fig. \ref{fig_ICs}. The left panel contains a meso-scale feature (the larger peak), and a sub-filter scale feature (the smaller peak). The right panel shows the same initial velocity but with added noise. The noise is a realization of a uniformly distributed random variable. In the sequel, we refer to the first initial condition as deterministic, while the second initial condition is called noisy.



\begin{figure}[h] \begin{center}
\includegraphics[height=1.5in,angle=0]{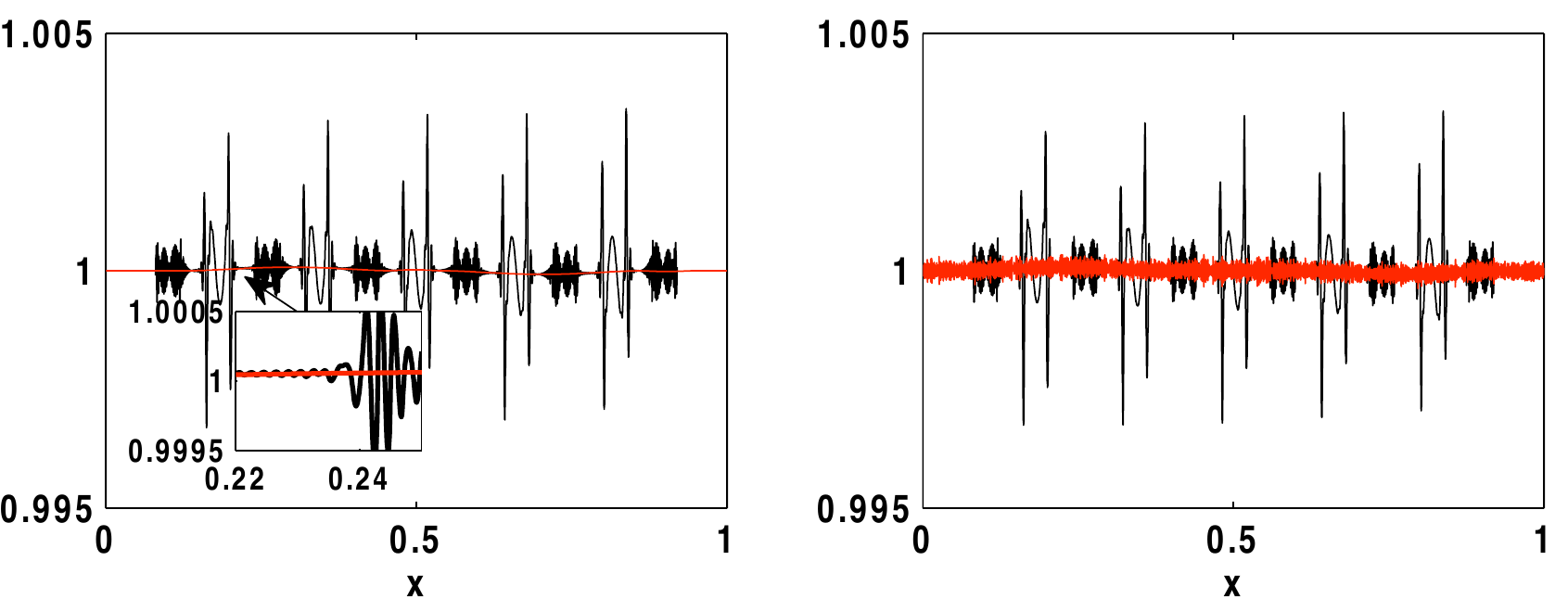}
\end{center}
\caption{Left panel: reconstruction of the Jacobian $J$ in the deterministic case; right panel: noisy case}
\label{fig-lj-jac}
\end{figure}

In Fig. \ref{fig-lj-jac}, we show exact and reconstructed Jacobians. The exact Jacobian in the deterministic case differs from the Jacobian in the noisy case, but the reconstructions are similar.  The similarity may be due to the built-in filtering in the deconvolution algorithm. This filtering is rather soft, since the reconstructed Jacobians contain oscillatory artifacts on scales comparable with the micro-scale. The amplitude of the artifacts is under control, so that the relative $l_\infty$ error does not exceed $0.3\%$. This shows that the reconstruction is stable.


\begin{figure}[h] \begin{center}
\includegraphics[height=1.5in,angle=0]{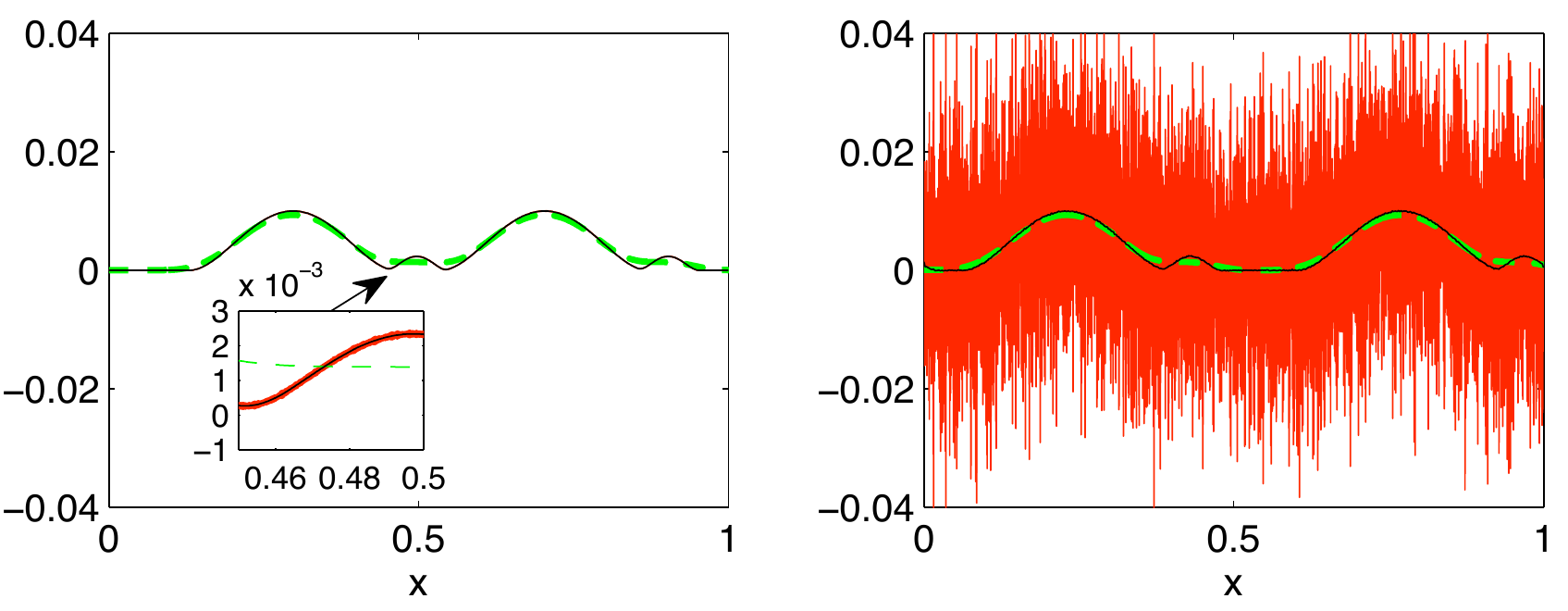}
\end{center}
\caption{Left panel: reconstruction of the velocity $\tilde v$ in the deterministic case; right panel: noisy case. On both panels the exact velocity is shown in red (dark grey) thin solid line, the average velocity in green (light grey) dashed line, and the reconstructed velocity in black solid line.
}
\label{fig-lj-vel}
\end{figure}

\begin{figure}[h] \begin{center}
\includegraphics[height=1.5in,angle=0]{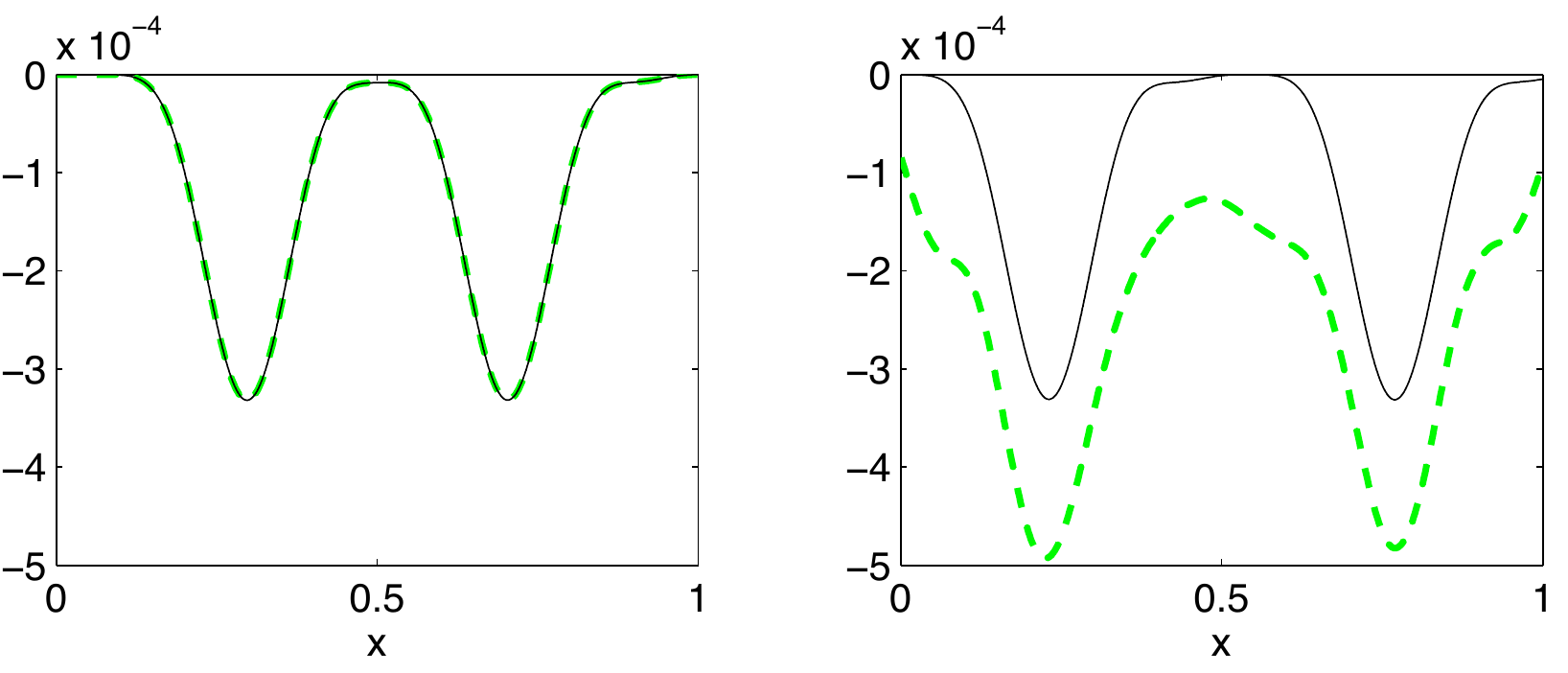}
\end{center}
\caption{Exact convective stress $T^\eta_{(c)}$ (green (grey) dashed line), and the approximation $\overline{T}^\eta_{(c)}$ (black solid line). Left panel:  deterministic case; right panel: noisy case. }
\label{fig-lj-conv}
\end{figure}

The velocity reconstruction is shown in Fig. \ref{fig-lj-vel}.  The average velocities in the noisy and deterministic case are nearly identical.  Averaging obliterates sub-lifter scale features, but the deconvolution algorithm recovers these features well. At the same time, the high frequency noise in the right panel is filtered out.
In Fig. \ref{fig-lj-conv}, we show the exact convective stress $T^\eta_{(c)}$ and its closed form approximation $\overline{T}^\eta_{(c)}$ computed from the equation
(\ref{c10}). We see that the approximation quality is better in the deterministic case.  In the noisy case, the main features of the stress are still well recovered, despite a significant difference between the exact and reconstructed velocities (Fig. \ref{fig-lj-vel}).

%

%

\begin{figure}[h] \begin{center}
\includegraphics[height=1.5in,angle=0]{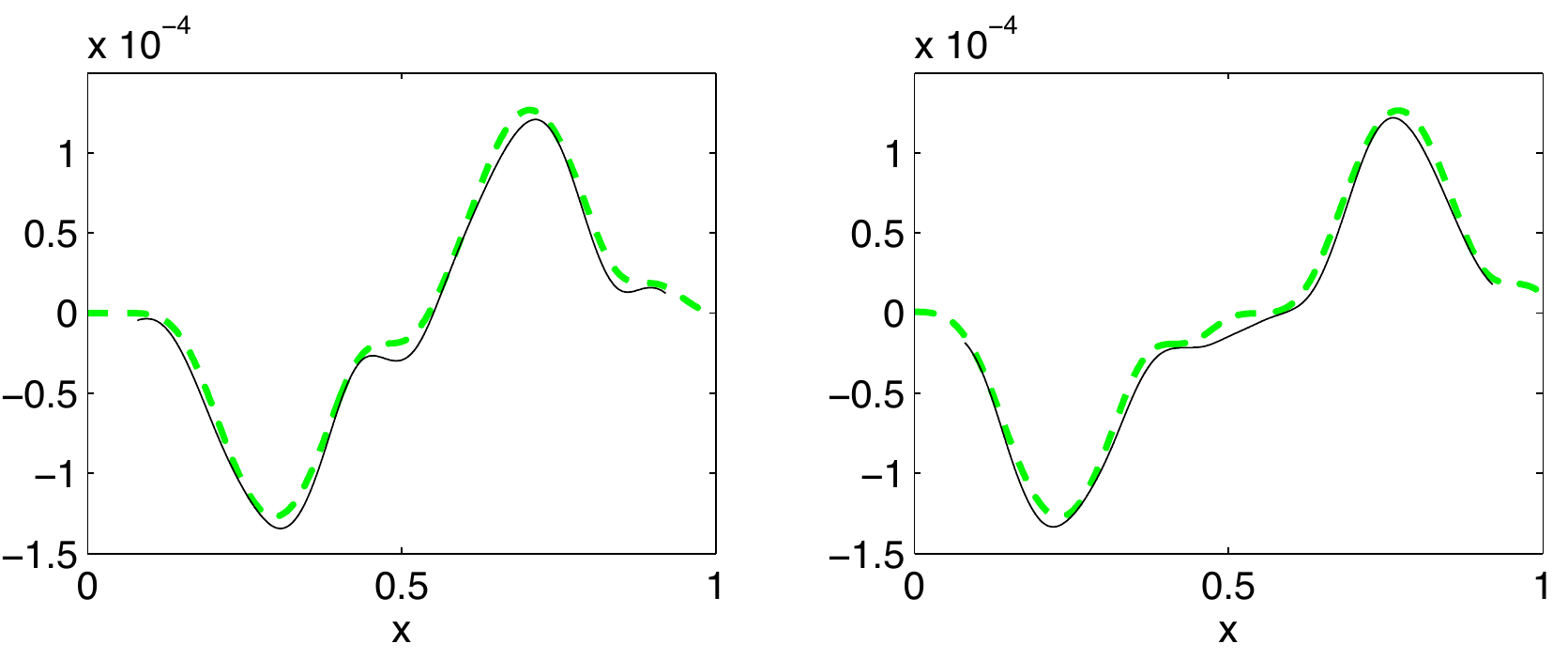}
\end{center}
\caption{Exact interaction stress $T^\eta_{(int)}$ (green (grey) dashed line), and the approximation $\overline{T}^\eta_{(c)}$ (black solid line). Left panel:  deterministic case; right panel: noisy case. }
\label{fig-lj-int}
\end{figure}

The interaction stress $T^\eta_{(int)}$ and its closed form approximation $\overline{T}^\eta_{(c)}$ (eq. (\ref{c11})) are shown in Fig. \ref{fig-lj-int}. The approximation quality is about the same in both cases. According to (\ref{c11}), the accuracy depends on the quality of the reconstruction of the Jacobian.  Comparison of Fig. \ref{fig-lj-jac} and Fig. \ref{fig-lj-int} shows that the high frequency artifacts and noise in the Jacobian are smoothed out by averaging in (\ref{c11}). This is important since the interaction stress depends non-linearly on the Jacobian.  In contrast to the linear case,  non-linear averaging functionals may exhibit sensitivity to oscillations in the input function (in this case, Jacobian).  Non-linearity induces dispersion, and this may result in transfer of the input's high frequency content into the low frequency content of the functional. In the present case,  this does not happen. We conjecture that for generic molecular potentials such as Lennard-Jones, and for a broad class of initial conditions, the stress functional has the self-averaging property, meaning that the effect of dispersion is weak compared to the filtering effect of convolution.
\subsection{Granular acoustics} \label{GAP}

\begin{figure}[h] \begin{center}
\includegraphics[height=1.5in,angle=0]{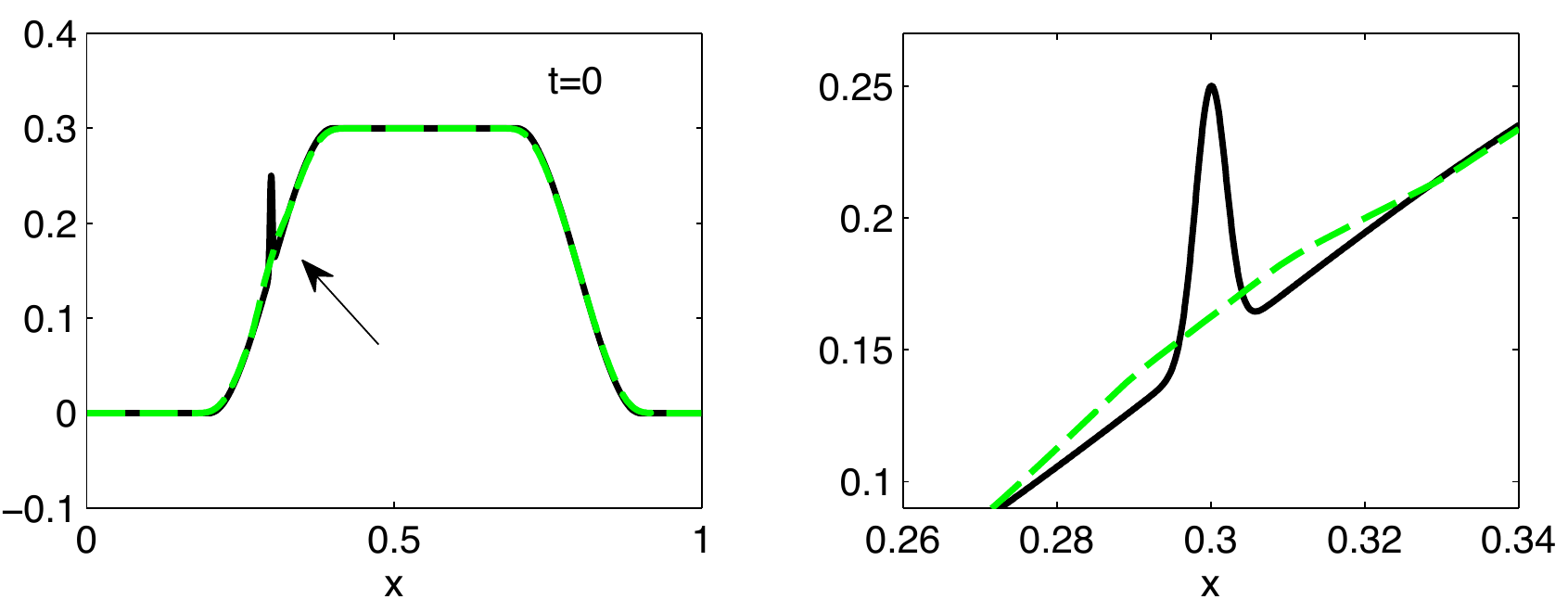}
\end{center}
\caption{Initial microscopic velocity $v^0$ (black solid curve) defined in (\ref{ICvel1}), (\ref{ICvel2}) together with the average velocity $\bar v$ (green dashed curve). Left panel: velocities are plotted on the entire domain $[0,1]$;  right panel: zoom-in of velocities on the interval $[0.26, 0.34]$ that contains a sub-filter feature}
\label{FLB0potential}
\end{figure}

In this subsection, we test the  method on a chain of particles interacting with  a pair potential $U(\xi)$ defined  in the Appendix in (\ref{gapdef}) and depicted in the right panel of Fig. \ref{FLB_Hertz_potential}.  The particles represent the centers of spherical granules and the potential resembles a Hertz potential employed in modeling of granular materials. The corresponding force  is purely repulsive and has a finite range equal to the equilibrium distance between the neighboring particles.

We solve the system of ODEs (\ref{ode-1}), (\ref{ode-2}) with two different initial conditions and periodic boundary conditions.
In both examples, the initial positions $q_j$ are equally spaced on the interval $(0,L)$ at the equilibrium distance $h=L/N$ with $L=1$ and $N=10,\!000$. The initial velocity for the first example, shown in Fig. \ref{FLB0potential} (black curve) and defined in the Appendix  (\ref{ICvel1}), (\ref{ICvel2}), contains features of different length scales.  The size of the larger feature of trapezoidal shape is bigger than $\eta L$ (mesoscale or filter scale). The smaller feature is a Gaussian with the standard deviation $0.2\eta L$ (sub-filter scale). At this length scale, the feature is completely obscured by the averaging (green dashed curve)  as can be seen in the right panel of Fig. \ref{FLB0potential} where the velocity is zoomed around $x=0.3$.

\begin{figure}[h] \begin{center}
\includegraphics[height=1.5in,angle=0]{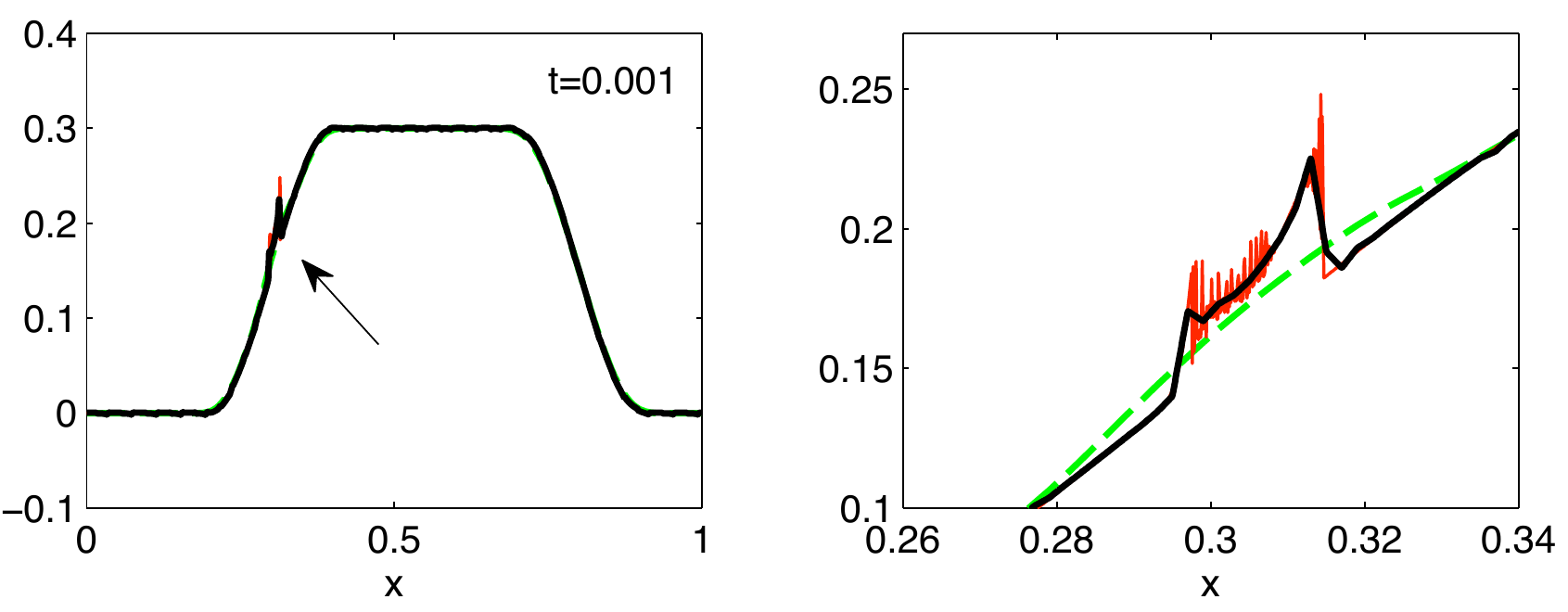}
\end{center}
\caption{Velocity reconstruction: microscopic velocity $\tilde v$ (red thin solid curve), reconstructed velocity $\frac{Q_\eta[\overline{\rho}^
\eta \overline{v}^\eta]}{Q_\eta [\overline{\rho}^
\eta]}$ (black thick solid curve) and average velocity $\bar v$ (green dashed curve). Left panel: on the entire domain; right panel: zoom-in of the region that contains a feature due to Gaussian perturbation}
\label{FLBvel1}
\end{figure}

The system of ODEs   (\ref{ode-1}), (\ref{ode-2}) is solved numerically until $t=2.2\cdot 10^{-2}$ after which the solution develops a shock. First, we test the quality of reconstruction of  $\tilde v$ and  $J$. Figure \ref{FLBvel1} compares the exact  $\tilde v$ (red thin solid curve), its reconstructed approximation $\frac{Q_\eta[\overline{\rho}^
\eta \overline{v}^\eta]}{Q_\eta [\overline{\rho}^
\eta]}$ (black thick solid curve) and the average  $\bar v$ (green dashed curve) at $t=10^{-3}$. The left panel shows that the reconstruction
captures the large scale features (left panel) as well as features on  the sub-filter scale  (right panel). In contrast, the average velocity completely misses the sub-filter scale.
%
%
%
\begin{figure}[h] \begin{center}
\includegraphics[height=1.5in,angle=0]{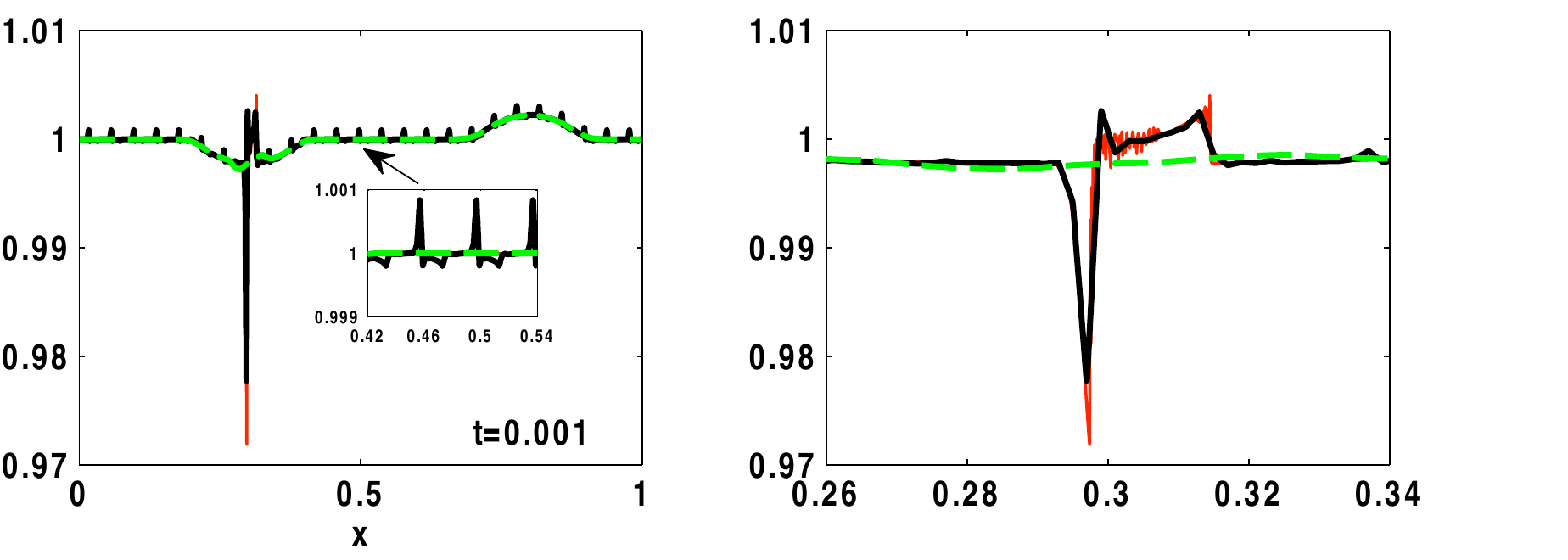}
\end{center}
\caption{Jacobian reconstruction: microscopic Jacobian $J$ (red thin solid curve), reconstructed Jacobian $\frac LM Q_\eta[\overline\rho^\eta]$ (black thick solid curve) and average  density $\frac LM\rho$ (green dashed curve). Left panel: solutions are shown on the entire domain; right panel: zoom-in of the region with sub-filter scale features}
\label{FLB2}
\end{figure}
In Fig. \ref{FLB2}, we compare the exact microscopic Jacobian $J$ (red thin solid curve) with the reconstructed Jacobian $\frac LM Q_\eta[\overline\rho^\eta]$ (black thick solid curve) and the average scaled density $\frac LM\rho$ (green dashed curve).
Similar to velocity reconstruction, Fig. \ref{FLB2} indicates that the  reconstructed Jacobian is much closer to the exact Jacobian than the average  density.



\begin{figure}[h] \begin{center}
\includegraphics[height=1.5in,angle=0]{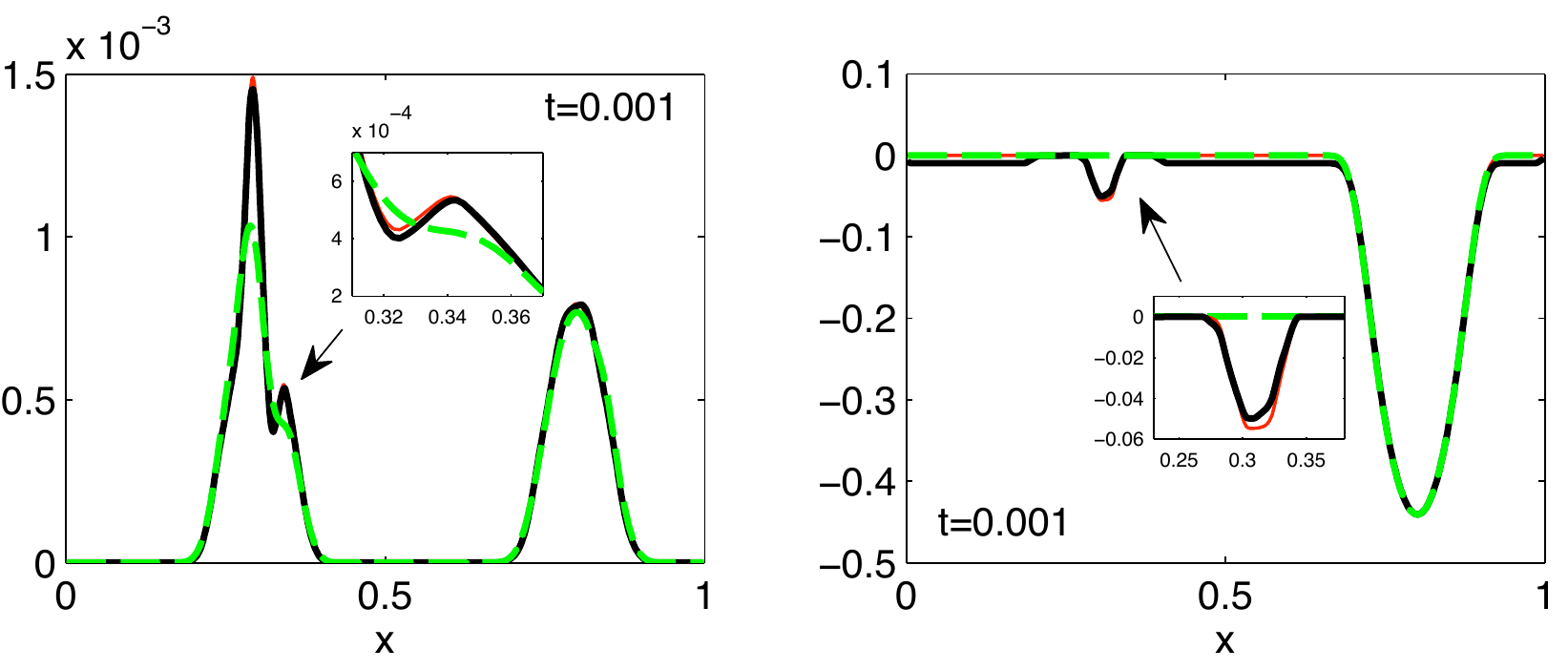}
\end{center}
\caption{Left panel shows convective stresses: exact $T^\eta_{(c)}$ (red thin solid curve), its approximation via reconstruction $\overline{T}^\eta_{(c)}$ (black thick solid curve), and an approximation (green dash curve) using zero-order closure (\ref{zero-order});
right panel: interaction stress $T^\eta_{(int)}$ and its corresponding approximations}
\label{FLBstresses}
\end{figure}

Next we examine how well  convective $T^\eta_{(c)}$ and  interaction  $T^\eta_{(int)}$ stresses are approximated by $\overline{T}^\eta_{(c)}$ and $\overline{T}^\eta_{(int)}$ defined in (\ref{c10}), (\ref{c11}). The left panel of Fig. \ref{FLBstresses} indicates that the exact convective stress and its approximation are almost indistinguishable.
The right panel of Fig. \ref{FLBstresses} shows good agreement between the exact interaction stress and its approximation.
For comparison, we also plot an approximation using a zero-order closure from \cite{PBG} shown for convenience in (\ref{zero-order}) that fails to capture sub-filter scale features in both stresses. 

The $l_\infty$-error in approximation of $T^\eta_{(c)}$ by $\overline{T}^\eta_{(c)}$ is between $1.5\%$ and $10\%$ during the simulation time. The error in approximation of $T^\eta_{(int)}$ is smaller and varies from $1.5\%$ to $8\%$. Preliminary computational studies indicate that the error decreases as $N$ increases.


%
\begin{figure}[h] \begin{center}
\includegraphics[width=4in,angle=0]{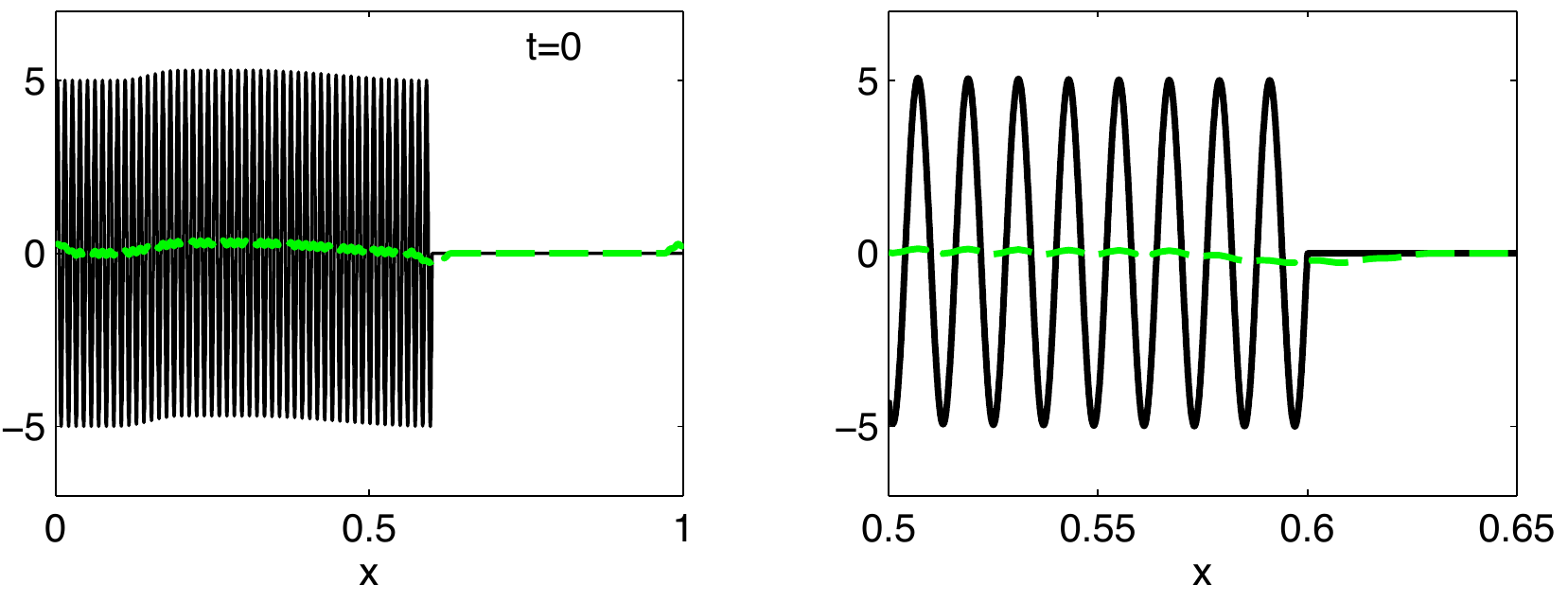}
\end{center}
\caption{Initial velocity with sine perturbation:  microscopic velocity $v^0$ (black solid curve) and average velocity $\bar v$ (green dashed curve). Left panel: velocity is shown on the entire domain; right panel: zoom-in of the velocity on the interval $[0.5, 0.65]$}
\label{FLBsine1}
\end{figure}

\begin{figure}[h] \begin{center}
\includegraphics[width=4in,angle=0]{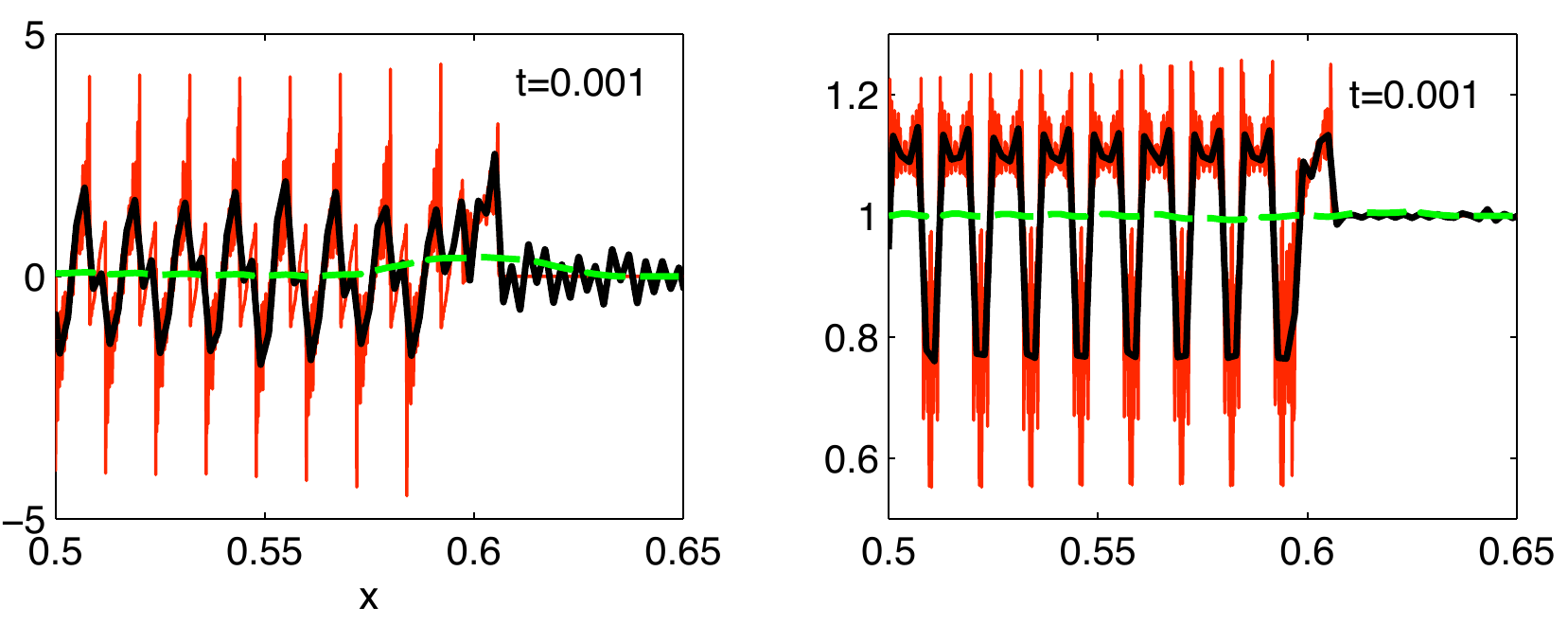}
\end{center}
\caption{Reconstruction of velocity $\tilde v$ (left panel); reconstruction of the Jacobian (let panel) at $t=10^{-3}$.
The functions are plotted on the interval $[0.5, 0.65]$ to show details. 
}
\label{FLBsine8}
\end{figure}

The initial velocity in the second example is the sum of $v^{base}$ defined in (\ref{ICvel1}) and a sine function with period $0.012$,  added on the interval $[0, 0.6]$ (see Fig. \ref{FLBsine1} and formula in (\ref{ICsine}) in the Appendix).
Simulations were also done until $t=2.2\cdot 10^{-2}$.  The right panel of Fig. \ref{FLBsine1}  shows that at $t=0$ the average velocity does not contain oscillations present in the microscopic velocity.
Fig. \ref{FLBsine8} presents graphs of velocity and Jacobian at a representative moment of time, $t=10^{-3}$.   The reconstructed velocity and Jacobian contain main features of their  microscopic counterparts while the averages do not.

%
%
%
%

\begin{figure}[h] \begin{center}
\includegraphics[width=4in,angle=0]{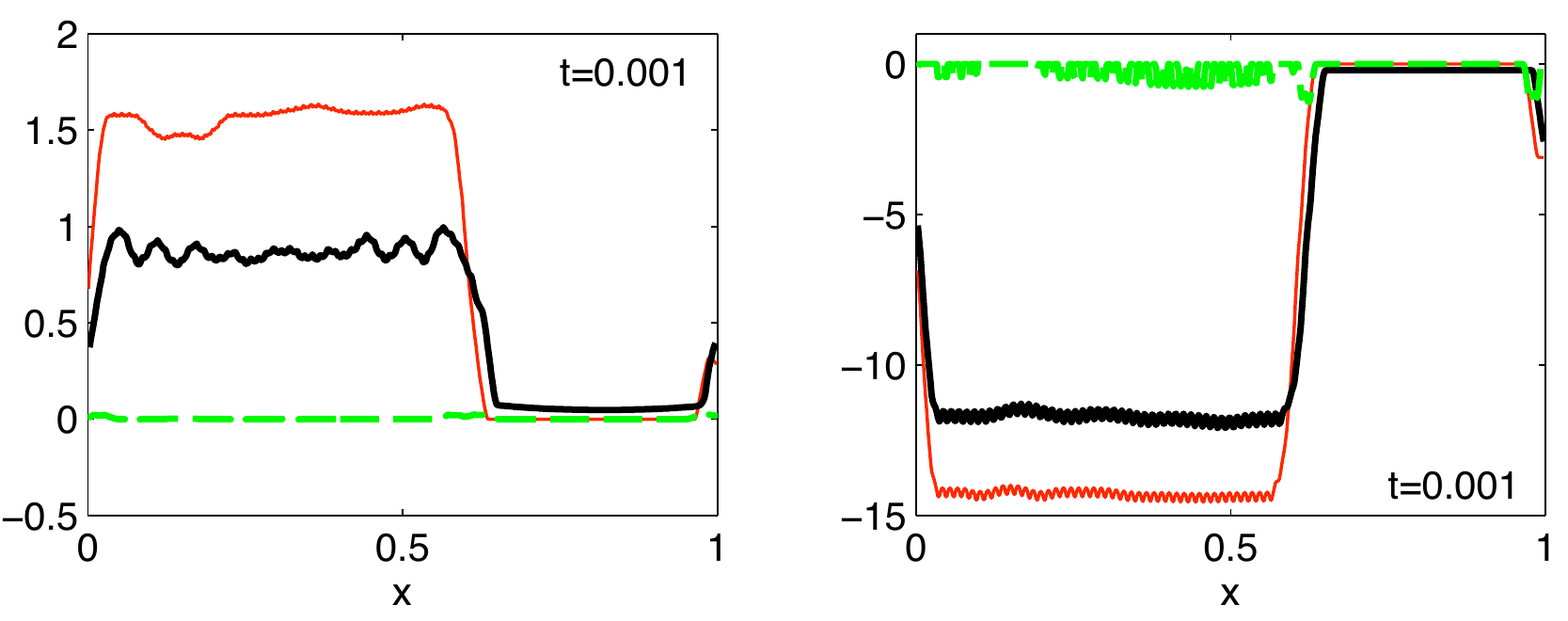}
\end{center}
\caption{Left panel: convective stress; right panel: interaction stress}
\label{FLBsine9}
\end{figure}

Fig. \ref{FLBsine9} depicts the  stress. Both convective and interaction exact stresses have sharp transition regions near $x=0.05$ and $x=0.6$. Our closed form approximation qualitatively captures these features
while a zero-order closure approximation is nearly zero on the entire interval.
The error in approximation of the convective stress fluctuates at early times (until $t=3\cdot 10^{-3}$) from $15\%$ to $70\%$ and then settles around $35-40\%$. The error in the approximation of the interaction stress behaves similarly at early times,
 then decreases to $10\%$ and levels off. The error in using the zero-order closure 
 is much higher: $75-100\%$ for the convective stress during the entire simulation time and around $100\%$ for the interaction stress until $t=2\cdot 10^{-3}$, then it drops to $10-15\%$ at $t=7\cdot 10^{-3}$, after which the error is about the same as using the reconstruction.

%



\section{Conclusions}
We propose a method for deriving closed form mesoscale continuum models of large particle systems. The closure construction is based on the following. Non-linear meso-scale averages can be rewritten as linear convolutions of the window function and appropriate micro-scale dynamical functions. One such function of particular importance is the inverse Jacobian of the micro-scale flow map associated with a position interpolant. Using the theory of ill-posed problems, we produce stable deconvolution approximations of particle positions and velocities in terms of the average density and average momentum.
Closure is achieved by inserting  these approximations into the equations for fluxes instead of the actual positions and velocities. The resulting constitutive equations (\ref{c10}), (\ref{c11}) are non-local in space and non-linear.

In the simplest version of the method, the micro-scale quantities are approximated by their averages.
We studied this approximation in the earlier paper \cite{PBG}. The results presented there indicate that the simplest approximation works well for systems characterized by (i) small fluctuations of the initial velocity; and (ii) nearly isothermal dynamics. In this article we consider more general initial conditions that contain prominent small scale peaks, significant noise, or high frequency periodic oscillations. Averaging obscures these features to such an extent that the approximation from \cite{PBG} becomes unsatisfactory. Here we were able to recover such details using non-iterative regularization methods.

We tested the method numerically on two models of FPU-chains: the classical Lennard-Jones chain, and the granular acoustics model considered earlier in \cite{PBG}, but with more general initial conditions. The ODEs were solved by the velocity Verlet method, and the obtained particle positions and velocities were used to calculate the average density, linear momentum, and the exact stress.  Then we used regularized deconvolution to generate the approximate (reconstructed) Jacobian and velocity. The resulting closed form approximation of the stress agreed very well with its exact counterpart.


\section{Acknowledgments}


Work of Lyudmyla Barannyk was supported in part by Amendment No. 005 to Task Order No. 00041 Under Master Task Agreement No. 00042246 Battelle Energy Alliance, LLC (BEA).


\appendix

\section{Window function}

\begin{figure}[h] \begin{center}
\includegraphics[height=1.5in,angle=0]{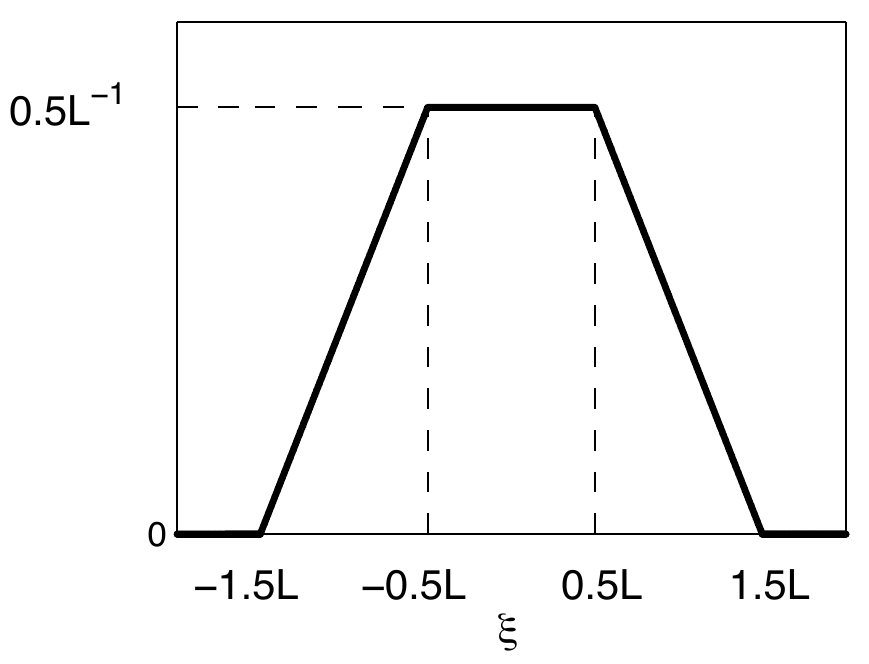}
\end{center}
\caption{The function $\psi(\xi)$}
\label{FLBpsi_window}
\end{figure}

In this paper, we use the following function $\psi$:
\begin{equation}\label{psi_window_def}
\psi(\xi)=
\left\{
\begin{array}{l}
\frac{1}{a+b}, \hspace{13pt}\quad \mbox{if} \quad |\xi|\leq a, \\[5pt]
\frac{\xi-b}{a^2-b^2}, \hspace{7pt} \quad \mbox{if} \quad a<\xi<b, \\[5pt]
-\frac{\xi+b}{a^2-b^2}, \quad \mbox{if} \quad -b<\xi<-a, \\[5pt]
0, \hspace{27pt} \quad \mbox{if} \quad |\xi|\geq b
\end{array}
\right.
\end{equation}
with $L=1$, $a=L/2$, $b=3L/2$. It can be directly checked that $\int_{-\infty}^\infty \psi(\xi) d\xi=\int_{-b}^{b} \psi(\xi) d\xi=1$. The function $\psi$ is plotted in Fig. \ref{FLBpsi_window}.

\section{Potentials and initial conditions}

In Section \ref{NumExper}, we test the method using two potentials: the first  is the Lennard-Jones potential, the second potential is similar to the Hertz potential employed in modeling of granular materials.

\subsection{Lennard-Jones}
\begin{figure}[h] \begin{center}
\includegraphics[height=1.5in,angle=0]{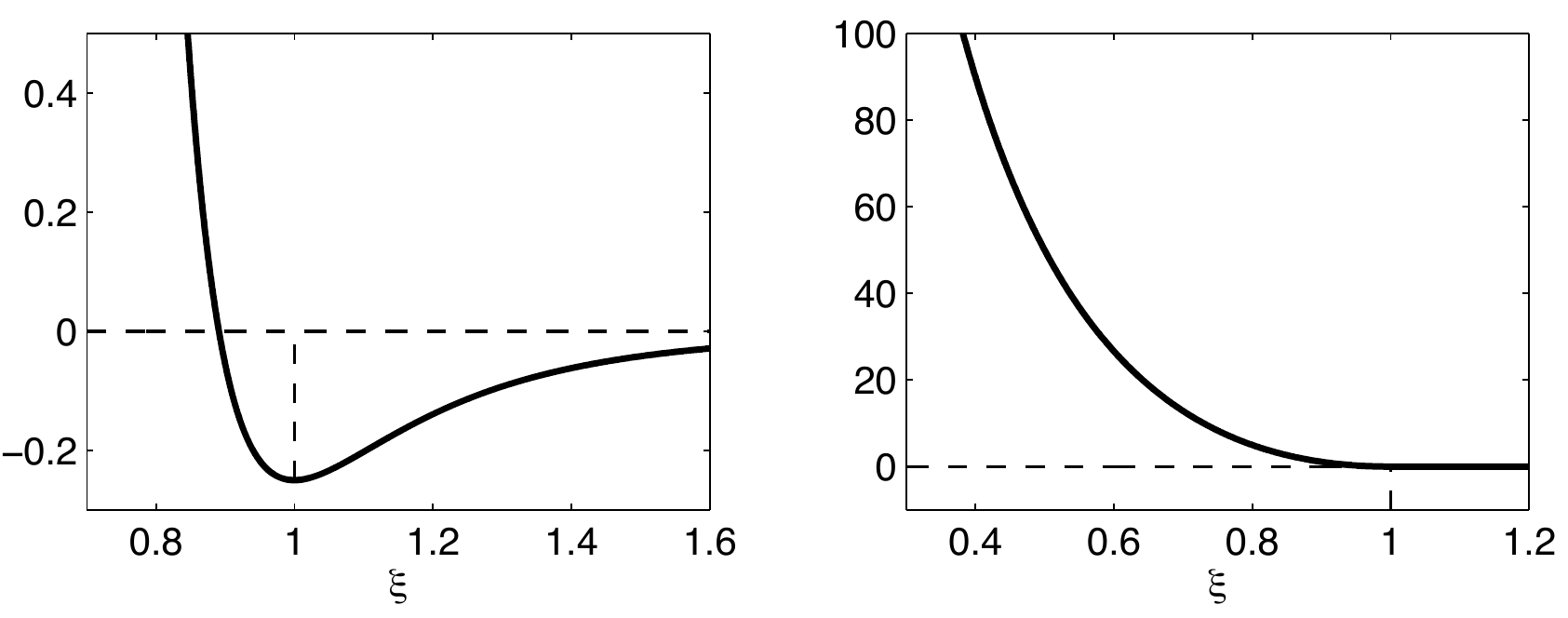}
\end{center}
\caption{Left panel: Lennard-Jones potential; right panel: Hertz potential}
\label{FLB_Hertz_potential}
\end{figure}

The potential is plotted in the left panel of Fig. \ref{FLB_Hertz_potential}
and defined by
\begin{equation}\label{LJ_def}
U(\xi) = 4\epsilon\left[\left(\frac{\sigma}{\xi}\right)^{12}-\left(\frac{\sigma}{\xi}\right)^{6}\right],
\end{equation}
where $\epsilon=0.25$  defines the depth of the potential well, $\sigma$ is the finite distance at which the potential is zero, $\xi$ is the distance between particles.
The potential is at a minimum when $\xi=h=2^{1/6}\sigma$, which determines the choice of $\sigma$.
 The force corresponding to the Lennard-Jones potential is repulsive for distances smaller than $h$ and attractive for distances greater than $h$.

The initial particles positions in all numerical test problems are equally spaced  with step $h$ and defined by $q_j^0=(j-1/2)h$, $j=1,\ldots,N$.
The initial velocity used in the deterministic case 
 is
%
%
%

%
\begin{equation}\label{IC1LJ}
v(q_j^0)= f(q_j^0)+\lambda\left(q_j^0-0.7\right), \quad   j=1,\ldots,N
\end{equation}
%
%
%
where
\[
f(\xi) =
\left\{
\begin{array}{l}
\frac{1}{50} \left(\xi-\frac 13\right)^2 \left(\frac 23-\xi\right)^2 \quad \mbox{if} \quad \frac 13<\xi<\frac 23, \\[5pt]
0 \quad \hspace{87pt} \mbox{otherwise}.
\end{array}
\right.
\]
and
$$
\lambda(\xi)=
\left\{
\begin{array}{l}
660\left(\frac{1}{20^2}-\xi^2\right)^2  \quad \text{if} \quad  -\frac{1}{20}< \xi < \frac{1}{20},
\\[5pt]
0 \quad \hspace{65pt} \mbox{otherwise}.\\
\end{array}
\right.
$$


%
%

The initial velocity for the noisy case is the sum of $v_0$ from (\ref{IC1LJ}) and a
uniformly distributed random variable with mean zero and maximum amplitude $10^{-3}$.

\subsection{Granular acoustics}
%
The potential is defined as
\begin{equation} \label{gapdef}
U(\xi)=
\left\{
\begin{array}{ll}
C_r\left(\frac{1}{1-p}\xi^{1-p}x_\star-\xi x_\star^{1-p}+\frac{p}{p-1}x_\star^{2-p}\right), & \quad {\rm if}~\xi\in (0, x_\star]\\
0,  & \quad {\rm if}~\xi > x_\star\\
\end{array}
\right.
\end{equation}
where $p>1$, $x_\star= L$, and $C_r$ is material stiffness. The potential is plotted in the right panel of Fig. \ref{FLB_Hertz_potential}.

The initial velocity in the first example of Subsection \ref{GAP} is given by $v^0=v^{base}+v^{1,pert}$, where $v^{base}$ is a piecewise cubic continuos function and $v^{1,pert}$ is a Gaussian:
%
\begin{eqnarray}\label{ICgauss}
 v^{base}(q_j^0)&=&
\left\{
\begin{array}{l}
0, \hspace{105pt}  \mbox{ if } \ 0\leq q_j^0 \leq L_1, \\[5pt]
d_1(q_j^0-x_1)(q_j^0-L_1)^2, \hspace{15pt} \mbox{ if } \ L_1< q_j^0 \leq L_2, \\[5pt]
d_2, \hspace{100pt}  \mbox{ if } \ L_2< q_j^0 \leq L_3, \qquad j=1,\ldots,N, \\[5pt]
d_3(q_j^0-x_2)(q_j^0-L_4)^2,  
\hspace{15pt}  \mbox{ if } \ L_3< q_j^0 \leq L_4, \\[5pt]
0, \hspace{105pt}  \mbox{ if } \ L_4< q_j^0 \leq L,
\end{array}
\right.
\label{ICvel1}
\\ [5pt]
 v^{1,pert}(q_j^0)&=&a_1\exp\left(-\frac{(q_j^0-q^*)^2}{2\sigma^2}\right),  \qquad j=1,\ldots,N.
\label{ICvel2}
\end{eqnarray}
%
Here $L_1=0.2L$, $L_2=0.4L$, $L_3=0.7L$, $L_4=0.9L$,
$x_1=({3L_2-L_1})/2$, $x_2=({3L_3-L_4})/2$,  $d_2=0.3$, $d_1=-{2d_2}/{(L_2-L_1)^3}$, $d_3=-{2d_2}/{(L_3-L_4)^3}$, $a_1=0.1$, $q^*=0.3L$.

The initial velocity in the second example is $v^0=v^{base}+v^{2,pert}$, where $v^{base}$ is as in (\ref{ICgauss}) but with $L_1=0.1L$, $L_2=0.2L$, $L_3=0.3L$, $L_4=0.6L$ and $v^{2,pert}$ is a sine function on the interval $[0, L_4]$:
\begin{equation}\label{ICsine}
v^{2,pert}(q_j^0)=
\left\{
\begin{array}{l}
a_2\sin\left(\frac{2\pi k q_j^0}{L_4}\right), \quad \mbox{if} \quad 0\leq q_j^0\leq L_4,  \qquad j=1,\ldots,N, \\[5pt]
0, \hspace{70pt} \mbox{otherwise}
\end{array}
\right.
\end{equation}
with $a_2=5$ and $k=50$. The sine perturbation has period $0.012$.

\bibliography{references_deconvolution}

\begin{thebibliography}{10}

\bibitem{Adams-Stolz2}
{\sc N.~A. Adams and S.~Stolz}, {\em A subgrid-scale deconvolution approach for
  shock capturing}, J. Comp. Phys., 178 (2002), pp.~391--426.

\bibitem{Layton}
{\sc L.~C. Berselli, T.~Iliescu, and W.~J. Layton}, {\em Mathematics of Large
  Eddy Simulation of Turbulent Flows}, Springer, New York, 2006.

\bibitem{Engl2}
{\sc H.~W. Engl, M.~Hanke, and A.~Neubauer}, {\em Regularization of Inverse
  Problems}, Kluwer Academic, Dordrecht, 1996.

\bibitem{Fridman}
{\sc V.~Fridman}, {\em A method of successive approximations for {F}redholm
  integral equations of the first kind}, Uspekhi Mat. Nauk, 11 (1956),
  pp.~233--234 (in Russian).

\bibitem{Gr}
{\sc C.~W. Groetsch}, {\em The Theory of {T}ikhonov Regularization for
  {F}redholm Equation of the First Kind}, Pitman, Boston, 1984.

\bibitem{Hansen}
{\sc P.~Ch. Hansen}, {\em Rank-Deficient and Discrete Ill-Posed Problems:
  Numerical Aspects of Linear Inversion}, SIAM, 1987.

\bibitem{Hardy}
{\sc R.~J. Hardy}, {\em Formulas for determining local properties in
  molecular-dynamics simulations: shock waves}, J. Chem. Phys., 76 (1982),
  pp.~622--628.

\bibitem{Kirkwood}
{\sc J.~H. Irving and J.~G. Kirkwood}, {\em The statistical theory of transport
  processes {IV}. {T}he equations of hydrodynamics}, J. Chem. Phys., 18 (1950),
  pp.~817--829.

\bibitem{Kirsch}
{\sc A.~Kirsch}, {\em An Introduction to the Mathematical Theory of Inverse
  Problems}, Springer, New York, 1996.

\bibitem{Land}
{\sc L.~Landweber}, {\em An iteration formula for {F}redholm integral equations
  of the first kind}, Am. J. Math., 73 (1951), pp.~615--624.

\bibitem{Layton07}
{\sc W.~Layton}, {\em Bounds on helicity and dissipation rates of approximate
  deconvolution models of turbulence}, SIAM J. Math. Anal., 39 (2007),
  pp.~916--931.

\bibitem{Layton06}
{\sc W.~Layton and O.~Lewandowski}, {\em Residual stress of approximate
  deconvolution models of turbulence}, J. Turbulence, 7 (2006), pp.~1--21.

\bibitem{Lehoucq2}
{\sc R.~B. Lehoucq and M.~P. Sears}, {\em The statistical mechanical foundation
  of the peridynamic nonlocal continuum theory: energy and momentum
  conservation laws}, Phys. Rev. E,  (to appear).

\bibitem{Morozov}
{\sc V.~A. Morozov}, {\em Methods for Solving Incorrectly Posed Problems},
  Springer, New York, 1984.

\bibitem{murdoch07}
{\sc A.~I. Murdoch}, {\em A critique of atomistic definitions of the stress
  tensor}, J. Elasticity, 88 (2007), pp.~113--140.

\bibitem{mb}
{\sc A.~I. Murdoch and D.~Bedeaux}, {\em Continuum equations of balance via
  weighted averages of microscopic quantities}, Proc. Royal Soc. London A, 445
  (1994), pp.~157--179.

\bibitem{mb96}
\leavevmode\vrule height 2pt depth -1.6pt width 23pt, {\em A microscopic
  perspective on the physical foundations of continuum mechanics -- {P}art {I}:
  macroscopic states, reproducibility, and macroscopic statistics, at
  prescribed scales of length and time}, Int. J. Engng Sci., 34 (1996),
  pp.~1111--1129.

\bibitem{mb97}
\leavevmode\vrule height 2pt depth -1.6pt width 23pt, {\em A microscopic
  perspective on the physical foundations of continuum mechanics -- {P}art
  {II}: a projection operator approach to the separation of reversible and
  irreversible contributions to macroscopic behaviour}, Int. J. Engng Sci., 35
  (1997), pp.~921--949.

\bibitem{Noll}
{\sc W.~Noll}, {\em Der herleitung der grundgleichungen der thermomechanik der
  kontinua aus der statistischen mechanik}, J. Ration. Mech. Anal., 4 (1955),
  pp.~627--646.

\bibitem{PBG}
{\sc A.~Panchenko, L.~L. Barannyk, and R.~P. Gilbert}, {\em Closure method for
  spatially averaged dynamics of particle chains}, Nonlinear Anal. Real World
  Appl., 12 (2011), pp.~1681--1697.

\bibitem{pavliotis-stuart}
{\sc G.~A. Pavliotis and A.~M. Stuart}, {\em Multiscale methods. Averaging and
  Homogenization}, Springer, 2008.

\bibitem{Lehoucq1}
{\sc S.~Silling and R.~B. Lehoucq}, {\em Peridynamic theory of solid
  mechanics}, Advances in Applied Mechanics, 44 (2010), pp.~73--168.

\bibitem{TPF}
{\sc A.~Tartakovsky, A.~Panchenko, and K.~Ferris}, {\em Dimension reduction
  method for {ODE} fluid models}, J. Comp. Phys.,  (to appear), p.~Preprint
  available at http://www.math.wsu.edu/math/faculty/panchenko/welcome.php.

\bibitem{Tikh1}
{\sc A.~N. Tikhonov and V.~Y. Arsenin}, {\em Solutions of Ill-Posed Problems},
  Wiley, New York, 1987.

\end{thebibliography}

\end{document}